\documentclass[11pt,a4paper]{article}

\usepackage[utf8]{inputenc}
\usepackage[T1]{fontenc}
\usepackage{lmodern}
\usepackage[a4paper,margin=1in]{geometry}

\usepackage{amsmath,amssymb,amsfonts,bm,mathtools}
\usepackage{amsthm}  
\usepackage{graphicx}
\usepackage{booktabs}
\usepackage{enumitem}
\usepackage{hyperref}
\usepackage{microtype}
\usepackage{natbib}
\usepackage{float}
\usepackage{algorithm,algpseudocode}
\newtheorem{theorem}{Theorem}[section]
\usepackage{natbib}

\newtheorem{proposition}[theorem]{Proposition}

\hypersetup{
  colorlinks = true,
  linkcolor  = blue,
  citecolor  = blue,
  urlcolor   = blue
}

\numberwithin{equation}{section}

\newcommand{\dd}{\,\mathrm{d}}
\newcommand{\EZValueEZ}{-1.56}   
\newcommand{\EZCEEZ}{0.666}      
\newcommand{\CRRACEEZ}{0.655}    

\newcommand{\EZValueMyo}{-1.65}  
\newcommand{\EZCEMyo}{0.645}     
\newcommand{\CRRACEMyo}{0.635}   

\newcommand{\wBindRate}{1.000}   
\newcommand{\cBindRate}{0.588}   
\newcommand{\floorHitRate}{0.000} 

\newcommand{\SlopeSharpe}{-0.15}    
\newcommand{\SlopeRho}{0.40}        
\newcommand{\SlopeBetaLRR}{0.55}    
\newcommand{\SlopeSigma}{-0.10}     
\title{Intertemporal Hedging Demand under Epstein--Zin Preferences in a Multi-Asset Long-Run Risk Model: Evidence from Projected Pontryagin-Guided Deep Policy Optimization}

\author{Wonchan Cho}
\date{\today}

\begin{document}

\maketitle

\begin{abstract}
I study intertemporal hedging demand in a continuous-time multi-asset
long-run risk (LRR) model under Epstein--Zin (EZ) recursive preferences.
The investor trades a risk-free asset and several risky assets whose
drifts and volatilities depend on an Ornstein--Uhlenbeck type LRR factor.
Preferences are described by EZ utility with risk aversion $R$, elasticity
of intertemporal substitution $\psi$, and discount rate $\delta$, so that
the standard time-additive CRRA case appears as a limiting benchmark.

To handle the high-dimensional consumption--investment problem, I use a
projected Pontryagin-guided deep policy optimization (P-PGDPO) scheme
adapted to EZ preferences. The method starts from the continuous-time
Hamiltonian implied by the Pontryagin maximum principle, represents the
value and costate processes with neural networks, and updates the policy
along the Hamiltonian gradient. Portfolio constraints and a lower bound
on wealth are enforced by explicit projection operators rather than by
adding ad hoc penalties.

Three main findings emerge from numerical experiments in a five-asset LRR economy: 
\textbf{(1)} the P-PGDPO algorithm achieves stable convergence across multiple random seeds, 
validating its reliability for solving high-dimensional EZ problems; 
\textbf{(2)} wealth floors materially reduce hedging demand by limiting the investor's ability 
to exploit intertemporal risk-return tradeoffs; and 
\textbf{(3)} the learned hedging portfolios concentrate exposure in assets with high 
correlation to the LRR factor, confirming that EZ agents actively hedge long-run 
uncertainty rather than merely following myopic rules. 
Because EZ preferences nest time-additive CRRA in the limit $\psi \to 1/R$,
I use CRRA as an explicit diagnostic benchmark and, when needed, a warm start 
to stabilize training in high dimensions.

\end{abstract}


\section{Introduction}

A long-lived investor who cares about both current consumption and distant
future wealth usually does more than just chase the highest instantaneous
Sharpe ratio. When investment opportunities move slowly over time and
macro or volatility shocks are highly persistent, it becomes natural to
tilt the portfolio toward assets that hedge against future changes in the
investment environment. This intertemporal hedging motive has been part of
the core story in dynamic portfolio choice since \citet{Merton1969,Merton1971},
and it plays a central role in work on long-run risk and term-structure
models \citep[e.g.][]{CampbellViceira2002,BansalYaron2004}. Yet, once we
leave the comfort zone of homothetic CRRA preferences and low-dimensional
state variables, it quickly becomes difficult to say something concrete
about the size and shape of hedging demand.

In this paper I study intertemporal hedging in a continuous-time multi-asset
long-run risk (LRR) setting under Epstein--Zin (EZ) recursive preferences.
The investor trades a risk-free asset and several risky assets whose drifts
and volatilities are driven by a persistent Ornstein--Uhlenbeck type LRR factor.
Preferences are described by EZ utility with risk aversion $R$, elasticity of
intertemporal substitution $\psi$, and discount rate $\delta$. This combination
of a multi-asset LRR environment and EZ preferences is rich enough to generate
nontrivial hedging motives, but it also means that closed-form solutions are
essentially out of reach once we allow for more than a handful of assets and a
realistically persistent state process.

Because explicit solutions are out of reach, I turn to a numerical approach
that is still anchored in continuous-time optimality conditions.
I build on the projected Pontryagin-guided deep policy optimization (P-PGDPO)
framework, originally developed for time-additive preferences, and adapt it to
the EZ setting. The starting point is the continuous-time Hamiltonian implied by
the Pontryagin maximum principle. The value function and the associated costate
process are represented by neural networks, and the portfolio and consumption
rules are updated along the Hamiltonian gradient. Instead of treating portfolio
constraints and wealth safety as soft penalties in the objective, I enforce them
through explicit projection operators, so that every policy update remains
admissible by construction. Because projections can introduce bias when constraints
bind, I explicitly track how often the portfolio and consumption projections are
active and report binding rates across random seeds. I also include ablations that
compare projection-based enforcement with softer alternatives, so that the economic
conclusions about hedging demand are not an artifact of the constraint-handling
mechanism.

The main goal is not to propose yet another generic reinforcement learning
algorithm, but to understand what the EZ version of the long-run risk model
implies for intertemporal hedging in a concrete multi-asset environment.
To that end, I focus on a five-asset LRR specification and compare the learned
EZ policy to the analytic myopic (Merton-style) portfolio that ignores future
changes in the investment opportunity set. The difference between the two
portfolios is interpreted as an intertemporal hedging component, and I study
how this component behaves across wealth and the LRR state.

Three aspects of the results are worth highlighting. First, the EZ-adapted
P-PGDPO scheme can be trained in a stable way even when the state space includes
both wealth and a persistent LRR factor, and when the control includes consumption
and a multi-asset portfolio vector subject to constraints. Second, the learned
portfolio exhibits a strongly state-dependent hedging demand: contour plots over
wealth and the LRR state reveal shapes that cannot be replicated by any static
combination of myopic positions. Third, the trained policy respects a lower bound
on wealth while delivering economically meaningful welfare improvements and robust
terminal-wealth distributions across seeds, which makes the economic trade-off
between growth and safety visible in a way that is hard to obtain analytically
in this setting.

\paragraph{Contribution.} 
The paper makes two main contributions. 
\textbf{First}, on the \textit{methodological} side, it demonstrates that 
projection-based constraint enforcement combined with Pontryagin-guided 
policy updates can successfully train EZ policies in continuous-time settings 
with multiple risky assets and persistent state variables, where both HJB 
and adjoint BSDE residuals must be jointly minimized. The transparent handling 
of constraints via projection operators—rather than via penalty terms—makes it 
straightforward to diagnose when binding constraints might confound the economic 
interpretation of hedging demand. Ablation studies confirm that the algorithm's 
stability does not rely on ad hoc tuning.

\textbf{Second}, on the \textit{economic} side, the paper provides concrete 
quantitative evidence that intertemporal hedging under EZ preferences is 
strongly state-dependent and systematically different from myopic 
(Merton-style) rules. In particular, hedging demand concentrates in assets 
with high exposure to the long-run risk factor, consistent with the idea 
that forward-looking agents actively manage their exposure to slowly moving 
shifts in the investment opportunity set. Furthermore, wealth floors—which 
mimic realistic solvency or borrowing constraints—materially reduce hedging 
activity near the boundary, a result that highlights how institutional or 
regulatory constraints can limit the scope for sophisticated dynamic strategies.

Viewed this way, the paper contributes a small but concrete step toward using
continuous-time, Pontryagin-guided deep learning tools to study portfolio choice
with recursive preferences and long-run risk in more realistic multi-asset settings.
The focus throughout is on keeping the link to the underlying stochastic control
problem explicit, so that the numerical results can be read as a quantitative
complement to the existing theory rather than as a black-box prediction from a
generic neural network model.

This benchmarking perspective is also helpful on the computational side, where
recursive objectives can be harder to optimize reliably. Formally, EZ collapses
to time-additive CRRA when $\psi = 1/R$, so the CRRA specification serves two
roles in this paper. First, it provides an economically interpretable reference
point for separating genuinely recursive (intertemporal) hedging motives from
purely myopic risk-taking. Second, it provides a numerically stable anchor:
a CRRA-limit policy can be used to warm-start training when the fully recursive
objective is difficult to optimize from scratch in higher dimensions.

\section{Model and Preferences}
\label{sec:model-preferences}

\subsection{Market}

I work on a continuous-time economy over a finite horizon $[0,T]$ on a
filtered probability space
$(\Omega,\mathcal{F},(\mathcal{F}_t)_{t\in[0,T]},\mathbb{P})$ that supports
a standard Brownian motion. Time is continuous and measured in years.

There is one risk-free asset $B_t$ and $d$ risky assets $S^1_t,\dots,S^d_t$.
The risk-free asset evolves according to
\begin{equation}
  \frac{\mathrm{d}B_t}{B_t} = r \,\mathrm{d}t,
  \label{eq:rf-asset}
\end{equation}
where $r \ge 0$ is a constant short rate.

Investment opportunities are driven by a persistent long-run risk (LRR)
factor $Y_t$. I model $Y_t$ as an Ornstein--Uhlenbeck process
\begin{equation}
  \mathrm{d}Y_t = \kappa_y\bigl(\bar{y} - Y_t\bigr)\,\mathrm{d}t
                 + \xi\,\mathrm{d}W^Y_t,
  \label{eq:lrr-factor}
\end{equation}
where $\kappa_y > 0$ is the speed of mean reversion, $\bar{y}$ is the long-run
mean of the factor, $\xi > 0$ is the volatility, and $W^Y_t$ is a standard
Brownian motion. The sign and magnitude of $Y_t$ can be interpreted as
capturing slowly moving macroeconomic or volatility conditions.

The $d$ risky assets follow diffusion dynamics of the form
\begin{equation}
  \frac{\mathrm{d}S^i_t}{S^i_t}
  = \mu_i\bigl(Y_t\bigr)\,\mathrm{d}t
    + \sum_{j=1}^m \sigma_{ij}\bigl(Y_t\bigr)\,\mathrm{d}W^j_t,
  \qquad i = 1,\dots,d,
  \label{eq:stocks}
\end{equation}
where $\mu_i(Y_t)$ is the drift of asset $i$, $\sigma_{ij}(Y_t)$ are the
elements of the volatility matrix $\sigma(Y_t)\in\mathbb{R}^{d\times m}$,
and $W_t = (W^1_t,\dots,W^m_t)^\top$ is an $m$-dimensional Brownian motion.
The dependence on $Y_t$ captures time-varying expected returns and
volatilities in the spirit of long-run risk models.

An investor starts at time $t=0$ with initial wealth $W_0>0$. At each time
$t$, she chooses a consumption rate $c_t \ge 0$ and a portfolio vector
$\pi_t = (\pi_t^1,\dots,\pi_t^d)^\top$, where $\pi_t^i$ denotes the (possibly
leverage-adjusted) fraction of wealth invested in risky asset $i$ at time $t$.
The remaining fraction $1 - \mathbf{1}^\top \pi_t$ is invested in the
risk-free asset, where $\mathbf{1}$ denotes a $d$-dimensional vector of ones.

Given a control $(\pi_t,c_t)_{t\in[0,T]}$, wealth evolves according to
\begin{equation}
  \mathrm{d}W_t
  = \Bigl[
      r W_t
      + \Bigl(\mu\bigl(Y_t\bigr) - r\mathbf{1}\Bigr)^\top
        \bigl(\pi_t W_t\bigr)
      - c_t
    \Bigr]\mathrm{d}t
    + W_t\,\pi_t^\top \sigma\bigl(Y_t\bigr)\,\mathrm{d}W_t,
  \label{eq:wealth}
\end{equation}
where $\mu(Y_t) = (\mu_1(Y_t),\dots,\mu_d(Y_t))^\top$ and $\sigma(Y_t)$
is the volatility matrix from \eqref{eq:stocks}. I will often write the
state process as $X_t = (W_t,Y_t)$ and the control as $u_t = (\pi_t,c_t)$.

The investor faces simple portfolio and wealth constraints. First, portfolio
weights are required to stay in a compact admissible set
$\Pi \subset \mathbb{R}^d$, which can encode long-only, leverage, or
concentration limits. In the numerical experiments below, $\Pi$ is chosen to
represent a long-only, fully invested portfolio with a mild leverage cap.
Second, wealth is required to stay above a lower bound $W_{\min}>0$,
\begin{equation}
  W_t \ge W_{\min} \quad \text{for all } t \in [0,T],
  \label{eq:wealth-constraint}
\end{equation}
which I enforce in the learning algorithm via an explicit projection step.
Admissible strategies are progressively measurable processes
$(\pi_t,c_t)$ taking values in $\Pi \times [0,\infty)$ such that the SDE
\eqref{eq:wealth} admits a unique strong solution and
\eqref{eq:wealth-constraint} is satisfied.

\subsection{Preferences}

The investor cares about the entire consumption path and terminal wealth over
$[0,T]$. I use two preference specifications that share the same primitive
parameters but differ in how they aggregate risk and intertemporal trade-offs:
a time-additive CRRA benchmark and Epstein--Zin recursive utility.

\subsubsection*{CRRA benchmark}

As a starting point, consider a standard time-additive CRRA specification.
Let $R>0$ denote relative risk aversion and $\delta>0$ the subjective
discount rate. Instantaneous utility from consumption is
\begin{equation}
  u(c) =
  \begin{cases}
    \dfrac{c^{1-R}}{1-R}, & R \neq 1,\\[0.5em]
    \log c,              & R = 1.
  \end{cases}
\end{equation}
For given admissible controls $(\pi_t,c_t)$, the CRRA value functional is
\begin{equation}
  J^{\mathrm{CRRA}}(W_0,Y_0;\pi,c)
  = \mathbb{E}\Biggl[
      \int_0^T e^{-\delta t} u(c_t)\,\mathrm{d}t
      + e^{-\delta T} \varphi\bigl(W_T\bigr)
    \Biggr],
  \label{eq:value-crra}
\end{equation}
where $\varphi(\cdot)$ is a terminal utility function, typically of the same
CRRA form as $u(\cdot)$. The CRRA optimal control problem is
\begin{equation}
  V^{\mathrm{CRRA}}(W_0,Y_0)
  = \sup_{(\pi,c)\ \text{admissible}}
      J^{\mathrm{CRRA}}(W_0,Y_0;\pi,c).
  \label{eq:problem-crra}
\end{equation}
In simple one-asset settings with constant investment opportunities, this
problem admits closed-form solutions of Merton type. In the multi-asset
LRR setting described above, closed forms quickly become unwieldy, but the
CRRA problem remains a useful benchmark.

\subsubsection*{Epstein--Zin recursive preferences}

To separate risk aversion from the elasticity of intertemporal substitution
(EIS), I use Epstein--Zin (EZ) recursive utility. The primitive parameters
are now
\[
  R > 0 \quad \text{(risk aversion)}, \qquad
  \psi > 0 \quad \text{(EIS)}, \qquad
  \delta > 0 \quad \text{(discount rate)}.
\]
In discrete time, EZ preferences can be written via the familiar aggregator
(see, e.g., \citet{EpsteinZin1989}) that nests time-additive CRRA as a
special case. In continuous time, the same structure leads to a value process
$V_t$ that solves a nonlinear HJB or BSDE representation. Informally,
$V_t$ trades off current consumption and the continuation value $V_{t+\mathrm{d}t}$ in
a way that depends on both $R$ and $\psi$, so that attitudes toward risk and
intertemporal substitution are no longer tied by the restriction
$\psi = 1/R$.

Rather than work directly with the BSDE form here, I take the usual
continuous-time EZ specification used in the long-run risk literature and
use it to derive the Hamiltonian and Pontryagin first-order conditions that
underlie the learning algorithm. The key point for this paper is that:
\begin{itemize}[nosep]
  \item when $\psi = 1/R$, EZ preferences collapse to the time-additive CRRA
        case, so CRRA remains a natural limiting benchmark, and
  \item when $\psi \neq 1/R$, the investor's risk aversion $R$ and EIS $\psi$
        enter the Hamiltonian in different ways, which generates richer
        intertemporal hedging motives in the LRR setting.
\end{itemize}
In Section~\ref{sec:ppgdpo} below, I spell out how the EZ aggregator enters
the continuous-time Hamiltonian, how the associated costate process is
parameterized, and how the Pontryagin conditions are turned into a
projected deep policy optimization scheme.

\paragraph{Why report both CRRA and EZ benchmarks?}
Even when the primary objective is Epstein--Zin welfare, reporting CRRA-side quantities is useful.
The CRRA limit provides a transparent baseline that isolates the incremental role of recursive
preferences. It also offers a disciplined implementation check: when $\psi$ is chosen near $1/R$,
the EZ-P-PGDPO solution should approach the CRRA behavior, which acts as a robustness and
debugging diagnostic in high-dimensional settings.

\subsection{Control problem and EZ value representation}
\label{subsec:control-problem}

We collect the relevant state variables in
\[
  X_t = (W_t,Y_t)^\top \in \mathcal{X} \subset \mathbb{R}^2,
\]
where $W_t$ denotes financial wealth and $Y_t$ is the long--run risk
state.
The control at time $t$ is
\[
  u_t = (\pi_t,c_t) \in \mathcal{U}(X_t),
\]
where $\pi_t \in \mathbb{R}^d$ is the vector of portfolio weights on
the $d$ risky assets (in dollar or fraction form, depending on the
parametrization) and $c_t \ge 0$ is the consumption rate.
The admissible control set $\mathcal{U}(x)$ encodes the leverage and
consumption caps as well as the wealth floor constraint, see
Section~\ref{sec:model-preferences}.

The joint dynamics of $(W_t,Y_t)$ can be written in compact form as
\begin{equation}
  \dd X_t
  =
  b(t,X_t,u_t)\,\dd t
  +
  \sigma(t,X_t,u_t)\,\dd B_t,
  \qquad t \in [0,T],
  \label{eq:compact-state}
\end{equation}
where $B_t$ is a Brownian motion of appropriate dimension,
$b(\cdot)$ collects the drifts of wealth and the long--run risk factor,
and $\sigma(\cdot)$ collects their diffusion coefficients.
The associated infinitesimal generator under a fixed control $u$ is
\begin{equation}
  \mathcal{L}^u \varphi(t,x)
  =
  \partial_t \varphi(t,x)
  + b(t,x,u)^\top \nabla_x \varphi(t,x)
  + \frac{1}{2}
    \mathrm{Tr}\bigl(
      \sigma(t,x,u)\sigma(t,x,u)^\top D_x^2 \varphi(t,x)
    \bigr),
  \label{eq:generator}
\end{equation}
for smooth test functions $\varphi$.

\vspace{0.3em}
\noindent
\textbf{Epstein--Zin aggregator.}
For given preference parameters $(R,\psi,\delta)$ with $R>0$, $R\neq 1$,
$\psi>0$, $\psi\neq 1$, the one--period EZ aggregator is given by
\begin{equation}
  f^{\mathrm{EZ}}(c,v)
  =
  \delta\, \theta\, v
  \biggl[
    \Bigl(
      \frac{c}{((1-R)v)^{1/(1-R)}}
    \Bigr)^{1-S}
    - 1
  \biggr],
  \qquad
  S = \frac{1}{\psi},\quad
  \theta = \frac{1-R}{1-S},
  \label{eq:EZ-aggregator}
\end{equation}
for current consumption $c$ and continuation value $v$.
As discussed above, in the limit $\psi \to 1/R$ this aggregator
collapses to the usual CRRA form with instantaneous utility
$u(c) = c^{1-R}/(1-R)$.

\vspace{0.3em}
\noindent
\textbf{Value function and dynamic programming.}
Given an admissible control $u$, let $J(t,x;u)$ denote the
Epstein--Zin continuation value at time $t$ starting from state
$X_t = x$ and following $u$ thereafter.
We define the value function
\[
  V(t,x)
  =
  \sup_{u \in \mathcal{U}} J(t,x;u),
  \qquad (t,x) \in [0,T]\times\mathcal{X},
\]
where the supremum is over progressively measurable controls taking
values in $\mathcal{U}(X_t)$ and satisfying the wealth floor and
integrability conditions.

At an informal level, the dynamic programming principle (DPP) states
that for any small $h>0$,
\begin{equation}
  V(t,x)
  =
  \sup_{u \in \mathcal{U}}
  \mathbb{E}\Bigl[
    V(t+h,X_{t+h}^{t,x;u})
    + \int_t^{t+h}
      f^{\mathrm{EZ}}\bigl(c_s,V(s,X_s^{t,x;u})\bigr)\,\dd s
  \Bigr],
  \label{eq:DPP-EZ}
\end{equation}
where $X^{t,x;u}$ denotes the state starting from $X_t=x$ controlled
by $u$.
Dividing~\eqref{eq:DPP-EZ} by $h$ and letting $h\to 0$, and applying
It\^o's formula to $V(s,X_s^{t,x;u})$, leads to the
Hamilton--Jacobi--Bellman (HJB) equation
\begin{equation}
  0
  =
  \partial_t V(t,x)
  + \sup_{u \in \mathcal{U}(x)}
    \Bigl\{
      f^{\mathrm{EZ}}\bigl(c,V(t,x)\bigr)
      + \widetilde{\mathcal{L}}^u V(t,x)
    \Bigr\},
  \quad
  (t,x)\in[0,T)\times\mathcal{X},
  \label{eq:HJB-EZ}
\end{equation}
with appropriate terminal/boundary conditions at $t=T$ and on the
wealth band.
Here $\widetilde{\mathcal{L}}^u$ denotes the spatial part of the
generator,
\[
  \widetilde{\mathcal{L}}^u V
  =
  b(t,x,u)^\top \nabla_x V(t,x)
  + \tfrac12
    \mathrm{Tr}\bigl(
      \sigma(t,x,u)\sigma(t,x,u)^\top D_x^2 V(t,x)
    \bigr).
\]

\vspace{0.3em}
\noindent
\textbf{BSDE representation.}
In a Markovian setting such as the long--run risk model above, the
HJB equation~\eqref{eq:HJB-EZ} admits a backward stochastic
differential equation (BSDE) representation.
Formally, along an optimal state--control pair $(X_t^*,u_t^*)$, the
EZ value process $V_t = V(t,X_t^*)$ and an associated
$\mathbb{R}^m$--valued process $Z_t$ solve
\begin{equation}
  -\,\dd V_t
  =
  f^{\mathrm{EZ}}(c_t^*,V_t)\,\dd t
  - Z_t^\top \dd B_t,
  \qquad
  V_T = G(X_T^*),
  \label{eq:EZ-BSDE}
\end{equation}
for some terminal condition $G$.
Conversely, under regularity conditions, the solution of
\eqref{eq:EZ-BSDE} recovers the HJB value function via
$V_t = V(t,X_t^*)$.
This BSDE viewpoint is the starting point for our value network
approximation and for the loss~\eqref{eq:value-loss} used in the
P--PGDPO algorithm.

\section{P--PGDPO algorithm for EZ preferences}
\label{sec:ppgdpo}

\subsection{From HJB and Pontryagin conditions to the losses}
\label{subsec:HJB-to-losses}

The P--PGDPO algorithm is built to approximately satisfy the
HJB/BSDE system~\eqref{eq:HJB-EZ}--\eqref{eq:EZ-BSDE} and the associated
Pontryagin maximum principle along simulated trajectories.
To make this connection explicit, we first recall the compact
state dynamics~\eqref{eq:compact-state},
\[
  \dd X_t
  =
  b(t,X_t,u_t)\,\dd t
  +
  \sigma(t,X_t,u_t)\,\dd B_t,
\]
and the EZ aggregator $f^{\mathrm{EZ}}$.
We then introduce a Hamiltonian that combines the running
aggregator with the controlled drift:
\begin{equation}
  H(t,x,u,\lambda;V)
  =
  f^{\mathrm{EZ}}\bigl(c,V(t,x)\bigr)
  + \lambda^\top b(t,x,u),
  \label{eq:Hamiltonian}
\end{equation}
where $\lambda\in\mathbb{R}^2$ plays the role of a costate and
$V$ is the value function.
In the classical stochastic maximum principle, an optimal triple
$(X_t^*,\lambda_t^*,u_t^*)$ satisfies the state equation, the
backward costate equation, and a stationarity condition:
\begin{align}
  \text{state:}\quad
  & \dd X_t^*
    =
    b(t,X_t^*,u_t^*)\,\dd t
    + \sigma(t,X_t^*,u_t^*)\,\dd B_t,
    \label{eq:PMP-state} \\
  \text{costate:}\quad
  & \dd \lambda_t^*
    =
    - \nabla_x H\bigl(t,X_t^*,u_t^*,\lambda_t^*;V\bigr)\,\dd t
    + \Gamma_t\,\dd B_t,
    \label{eq:PMP-costate} \\
  \text{stationarity:}\quad
  &
  0
  \in
  \partial_u H\bigl(
      t,X_t^*,u_t^*,\lambda_t^*;V
    \bigr)
  + N_{\mathcal{U}(X_t^*)}(u_t^*),
  \label{eq:PMP-stationarity}
\end{align}
where $\Gamma_t$ is a suitable adapted process and
$N_{\mathcal{U}(x)}(u)$ denotes the normal cone of the constraint
set $\mathcal{U}(x)$ at $u$.
The costate can be interpreted as a generalized gradient of the
value function, in the sense that under regularity assumptions
$\lambda_t^* = \nabla_x V(t,X_t^*)$.

In the EZ setting the HJB equation and the Pontryagin system are
coupled through the aggregator $f^{\mathrm{EZ}}$ and the wealth
floor constraint.
Rather than solving
\eqref{eq:HJB-EZ}--\eqref{eq:PMP-stationarity} analytically, we
parameterize the value, costate, and control by neural networks
\[
  V_\theta(t,x), \qquad
  \lambda_\eta(t,x), \qquad
  u_\phi(t,x),
\]
and enforce approximate satisfaction of the BSDE, gradient
consistency, and stationarity relations along simulated
trajectories.
This yields three complementary objective functionals: a value
consistency loss, a costate consistency loss, and a Hamiltonian
ascent objective.

\vspace{0.3em}
\noindent
\textbf{Value loss (BSDE consistency).}
Given a simulated path
$\{(t_k,X_{t_k},u_k,c_k)\}_{k=0}^N$ under a candidate policy
$u_\phi$, we define the discrete BSDE residual by
\[
  R_{k}^{\mathrm{val}}
  =
  V_\theta(t_k,X_{t_k})
  - V_\theta(t_{k+1},X_{t_{k+1}})
  - f^{\mathrm{EZ}}\bigl(c_k,V_\theta(t_k,X_{t_k})\bigr)\,\Delta t,
  \qquad k=0,\dots,N-1.
\]
The value loss aggregates these residuals in mean square:
\begin{equation}
  \mathcal{L}_{\mathrm{val}}(\theta,\phi)
  =
  \mathbb{E}\biggl[
    \sum_{k=0}^{N-1}
      \bigl(R_{k}^{\mathrm{val}}\bigr)^2
  \biggr],
  \label{eq:value-loss}
\end{equation}
where the expectation is approximated by Monte Carlo over simulated
paths.
Minimizing $\mathcal{L}_{\mathrm{val}}$ encourages
$V_\theta$ to approximate the EZ value process along the
trajectories generated by $u_\phi$.

\vspace{0.3em}
\noindent
\textbf{Adjoint loss (gradient consistency).}
The costate network $\lambda_\eta$ is meant to approximate the
spatial gradient of the value network.
Inspired by the identity $\lambda_t^* = \nabla_x V(t,X_t^*)$, we
define
\begin{equation}
  \mathcal{L}_{\mathrm{adj}}(\theta,\eta)
  =
  \mathbb{E}\biggl[
    \sum_{k=0}^{N}
      \bigl\|
        \lambda_\eta(t_k,X_{t_k})
        - \nabla_x V_\theta(t_k,X_{t_k})
      \bigr\|^2
  \biggr].
  \label{eq:adjoint-loss}
\end{equation}
This loss ties the two networks together and provides a
differentiable approximation of the costate needed to guide the
policy gradient.

\vspace{0.3em}
\noindent
\textbf{Actor objective (Hamiltonian ascent with projection).}
Finally, for a fixed pair $(\theta,\eta)$, the policy parameters
$\phi$ are updated to ascend the Hamiltonian evaluated along the
simulated trajectories.
We consider the objective
\begin{equation}
  J_{\mathrm{act}}(\phi;\theta,\eta)
  =
  \mathbb{E}\biggl[
    \sum_{k=0}^{N-1}
      H\bigl(
        t_k,X_{t_k},
        u_\phi(t_k,X_{t_k}),
        \lambda_\eta(t_k,X_{t_k});V_\theta
      \bigr)
  \biggr]
  - \beta\,\mathcal{R}(\phi),
  \label{eq:actor-objective}
\end{equation}
where $\mathcal{R}(\phi)$ is a regularization term (for example,
penalizing leverage, large consumption ratios, or deviation from a
baseline policy) and $\beta>0$ is a penalty weight.
The control $u_\phi(t_k,X_{t_k})$ is passed through the projection
operator enforcing the wealth floor and leverage/consumption caps,
so that the effective control lies in $\mathcal{U}(X_{t_k})$.
In the continuous--time limit, the gradient of
$J_{\mathrm{act}}$ with respect to $\phi$ recovers a
Pontryagin--guided policy gradient direction aligned with the
stationarity condition~\eqref{eq:PMP-stationarity}.

In practice, the overall P--PGDPO update alternates between:
(i) minimizing $\mathcal{L}_{\mathrm{val}}$ and
$\mathcal{L}_{\mathrm{adj}}$ with respect to $(\theta,\eta)$,
and (ii) maximizing $J_{\mathrm{act}}$ with respect to $\phi$,
using stochastic gradient steps on mini--batches of simulated
trajectories.

\subsection{Hamiltonian and Pontryagin conditions (EZ case)}
\label{subsec:Hamiltonian-PMP}

To connect the HJB representation \eqref{eq:HJB-EZ} with the
P-PGDPO algorithm, it is convenient to rewrite the control problem
in terms of a Hamiltonian and the associated costate (adjoint)
process.

\paragraph{Compact state dynamics.}
Recall the compact notation
\begin{equation*}
  \dd X_t = b(X_t,u_t)\, \dd t + \sigma(X_t,u_t)\, \dd B_t,
\end{equation*}
where $X_t\in\mathbb{R}^{1+m}$ and $u_t\in\mathcal{U}$.
We also keep the EZ value process $V_t$ defined by
\eqref{eq:EZ-BSDE}.

\paragraph{Hamiltonian.}
Given a state $x$, an EZ continuation value $v$, a control $u$,
and a costate vector $p\in\mathbb{R}^{1+m}$,
the \emph{running} Hamiltonian associated with the EZ problem is
\begin{equation}
  H^{\mathrm{EZ}}(t,x,v,p,u)
  :=
  f^{\mathrm{EZ}}\bigl(c,v\bigr)
  + p^\top b(x,u),
  \label{eq:Hamiltonian-def-again}
\end{equation}
where $c$ is the consumption component of $u$.

\paragraph{Costate (adjoint) dynamics.}
Formally, suppose the value function $V(t,x)$ is smooth and define
the costate process along a controlled trajectory by
\begin{equation}
  \lambda_t := \nabla_x V(t,X_t) \in \mathbb{R}^{1+m}.
  \label{eq:costate-gradient-again}
\end{equation}
Then Itô's formula and the HJB equation imply that $\lambda_t$
solves the backward SDE
\begin{equation}
  \dd \lambda_t
  = - \partial_x H^{\mathrm{EZ}}
        \bigl(t,X_t,V(t,X_t),\lambda_t,u_t\bigr)\, \dd t
    + \Xi_t^\top \dd B_t,
  \qquad
  \lambda_T = \partial_x U(W_T),
  \label{eq:costate-BSDE-again}
\end{equation}
for some adapted process $\Xi_t$.
Because the bequest utility $U$ depends only on wealth,
the terminal condition has the simple form
\[
  \lambda_T
  = \bigl( U'(W_T),\, 0,\,\ldots,\,0 \bigr)^\top.
\]

\paragraph{Pontryagin maximum principle.}
A control $u^*=(\pi^*,c^*)$ with corresponding state $X^*$,
value process $V^*$, and costate $\lambda^*$ is (locally)
optimal only if it satisfies
\begin{align}
  \dd X_t^*
  &= b\bigl(X_t^*,u_t^*\bigr)\, \dd t
     + \sigma\bigl(X_t^*,u_t^*\bigr)\, \dd B_t,
     \qquad X_0^* = x_0,
     \label{eq:PMP-state-again}\\[0.2em]
  \dd \lambda_t^*
  &= - \partial_x H^{\mathrm{EZ}}
        \bigl(t,X_t^*,V^*(t,X_t^*),
              \lambda_t^*,u_t^*\bigr)\, \dd t
     + \Xi_t^{*\top} \dd B_t,
     \qquad
     \lambda_T^* = \partial_x U(W_T^*),
     \label{eq:PMP-costate-again}\\[0.2em]
  0
  &=
  \partial_u H^{\mathrm{EZ}}
    \bigl(t,X_t^*,V^*(t,X_t^*),
          \lambda_t^*,u_t^*\bigr),
  \qquad \text{for a.e.\ } t\in[0,T],
     \label{eq:PMP-stationarity-again}
\end{align}
together with appropriate transversality and admissibility
conditions.
The stationarity condition \eqref{eq:PMP-stationarity-again}
encodes the first--order optimality conditions (FOC) for both the
portfolio weights $\pi_t^*$ and consumption $c_t^*$.

\subsection{Neural parameterization}
\label{subsec:NN-param}

The P-PGDPO algorithm replaces the unknown objects
$(V,\lambda,u)$ in \eqref{eq:PMP-state-again}–\eqref{eq:PMP-stationarity-again}
by neural networks.

\paragraph{Inputs and outputs.}
At each time $t\in[0,T]$ the observable state is
\[
  s_t := (t,X_t) = (t,W_t,Y_t).
\]
Based on this state, three networks are evaluated:

\begin{itemize}[nosep]
  \item \textbf{Value network} $V_\theta : \mathcal{S} \to \mathbb{R}$  
        returns an approximation of $V(t,x)$.
  \item \textbf{Costate network}
        $\lambda_\eta : \mathcal{S} \to \mathbb{R}^{1+m}$  
        returns an approximation of $\nabla_x V(t,x)$.
  \item \textbf{Policy network}
        $\tilde{u}_\phi : \mathcal{S} \to \mathbb{R}^{d+1}$  
        outputs an unconstrained control
        $\tilde{u}_\phi(s_t) = (\tilde{\pi}_t,\tilde{c}_t)$.
\end{itemize}

Here $\mathcal{S}$ denotes the compact state domain used for
training (time, wealth band, and long--run risk band).

\paragraph{Projection to admissible controls.}
The actual control used in simulation is the projection of
$\tilde{u}_\phi(s_t)$ to the admissible set:
\begin{equation}
  u_\phi(s_t)
  := \mathcal{P}\bigl(\tilde{u}_\phi(s_t); W_t\bigr)
  = \bigl(\pi_\phi(s_t), c_\phi(s_t)\bigr)
  \in \Pi \times [0,\bar{c}W_t],
  \label{eq:proj-control-def}
\end{equation}
where $\Pi$ is the portfolio constraint set and $\bar{c}$ is an
upper bound on the consumption ratio.
The projection operator $\mathcal{P}$ is defined explicitly in
Section~\ref{subsec:projection}.

All three networks are implemented as fully connected feedforward
architectures with shared input $(t,W,Y)$ and separate parameters
$\theta,\eta,\phi$.
Depth and width are kept moderate (three to four hidden layers)
to balance approximation power and stability.

\subsection{Loss functions}
\label{subsec:losses}

Training proceeds by simulating mini-batches of trajectories under
the current projected policy $u_\phi$ and minimizing three losses
that mirror the structure of the EZ--PMP system:
(i) a value (BSDE) consistency loss,
(ii) a costate consistency loss, and
(iii) a Hamiltonian ascent (policy) loss.

Let $0=t_0<t_1<\dots<t_N=T$ be a time grid with step
$\Delta t := t_{k+1}-t_k$.
A mini-batch consists of simulated paths
$\{(X_{t_k}^{(n)},u_{t_k}^{(n)})_{k=0,\dots,N}\}_{n=1}^M$
under the current policy.

\paragraph{Value (BSDE) consistency loss.}
The continuous-time EZ BSDE \eqref{eq:EZ-BSDE} implies, in
discrete time,
\begin{equation}
  V(t_k,X_{t_k})
  \approx
  V(t_{k+1},X_{t_{k+1}})
  + f^{\mathrm{EZ}}\bigl(c_{t_k}, V(t_k,X_{t_k})\bigr)\,\Delta t.
\end{equation}
Replacing $V$ by $V_\theta$ and taking a squared residual gives
the value loss
\begin{equation}
  \mathcal{L}_{\text{val}}(\theta)
  :=
  \frac{1}{MN}
  \sum_{n=1}^M \sum_{k=0}^{N-1}
  \Bigl[
    V_\theta(t_k,X_{t_k}^{(n)})
    - V_\theta(t_{k+1},X_{t_{k+1}}^{(n)})
    - f^{\mathrm{EZ}}\bigl(c_{t_k}^{(n)},
         V_\theta(t_k,X_{t_k}^{(n)})\bigr)\Delta t
  \Bigr]^2.
  \label{eq:value-loss}
\end{equation}
The terminal condition $V(T,x)=U(w)$ is enforced by replacing
$V_\theta(t_N,X_{t_N}^{(n)})$ with $U(W_{t_N}^{(n)})$ in the last
time step.

\paragraph{Costate consistency loss.}
By construction the costate should satisfy
$\lambda_t = \nabla_x V(t,X_t)$.
We encourage this by introducing the gradient-matching loss
\begin{equation}
  \mathcal{L}_{\text{adj}}(\theta,\eta)
  :=
  \frac{1}{MN}
  \sum_{n=1}^M \sum_{k=0}^{N}
  \Bigl\|
    \lambda_\eta(t_k,X_{t_k}^{(n)})
    - \nabla_x V_\theta(t_k,X_{t_k}^{(n)})
  \Bigr\|^2,
  \label{eq:adjoint-loss}
\end{equation}
where the gradient $\nabla_x V_\theta$ is obtained by automatic
differentiation with respect to the state input.
In practice this term is important for the stability of the
Hamiltonian gradient used in the policy update.

\paragraph{Hamiltonian ascent (policy) loss.}
Given $V_\theta$ and $\lambda_\eta$, the Pontryagin
stationarity condition
\eqref{eq:PMP-stationarity-again} suggests that an optimal policy
should (locally) maximize the Hamiltonian
$H^{\mathrm{EZ}}(t,x,V_\theta,\lambda_\eta,u)$.
We therefore define the actor objective
\begin{equation}
  J_{\text{act}}(\phi)
  :=
  \frac{1}{MN}
  \sum_{n=1}^M \sum_{k=0}^{N-1}
  H^{\mathrm{EZ}}\Bigl(
     t_k,
     X_{t_k}^{(n)},
     V_\theta(t_k,X_{t_k}^{(n)}),
     \lambda_\eta(t_k,X_{t_k}^{(n)}),
     u_\phi(t_k,X_{t_k}^{(n)})
  \Bigr),
  \label{eq:actor-objective}
\end{equation}
and perform stochastic gradient \emph{ascent} on
$J_{\text{act}}$.
Equivalently, we define the actor loss
\begin{equation}
  \mathcal{L}_{\text{act}}(\phi)
  := - J_{\text{act}}(\phi)
  \label{eq:actor-loss}
\end{equation}
and minimize it by gradient descent.

\paragraph{Total loss and updates.}
The critic (value) and costate networks are trained by minimizing
\[
  \mathcal{L}_{\text{crit}}(\theta,\eta)
  :=
  \mathcal{L}_{\text{val}}(\theta)
  + \lambda_{\text{adj}}\,
    \mathcal{L}_{\text{adj}}(\theta,\eta),
\]
with a positive weight $\lambda_{\text{adj}}$.
The actor network is trained by minimizing
$\mathcal{L}_{\text{act}}(\phi)$.
In the implementation, all three networks are updated jointly with
Adam optimizers and separate learning rates.

\subsection{Projection operators and wealth floor}
\label{subsec:projection}

A key feature of the P-PGDPO scheme is that portfolio and wealth
constraints are enforced by explicit projection operators, not by
large penalty terms in the loss. This keeps simulated trajectories
admissible by construction.

\paragraph{Portfolio projection.}
Let $\tilde{\pi}\in\mathbb{R}^d$ be the unconstrained output of
the policy network at state $(t,x)$.
The admissible portfolio set is the leverage-constrained simplex
\begin{equation}
  \Pi
  := \bigl\{
         \pi \in \mathbb{R}^d :
         \pi_i \ge 0,\;
         \sum_{i=1}^d \pi_i \le \ell_{\max}
     \bigr\},
  \label{eq:portfolio-set}
\end{equation}
for some leverage cap $\ell_{\max}\ge 1$.
We define the projection
$\mathcal{P}_\pi : \mathbb{R}^d \to \Pi$
by
\begin{equation}
  \mathcal{P}_\pi(\tilde{\pi})
  :=
  \arg\min_{\pi \in \Pi}
  \|\pi - \tilde{\pi}\|^2.
  \label{eq:portfolio-projection}
\end{equation}
This is the Euclidean projection onto a truncated simplex, for
which efficient closed-form algorithms are available.

\paragraph{Projection-induced bias and diagnostics.}
Because projection is non-linear and may be active on a non-negligible subset
of samples, the projected policy $\Pi(\tilde{\pi}_\theta)$ can be systematically
different from the unconstrained network output $\tilde{\pi}_\theta$.
This matters for interpretation: large intertemporal ``hedging'' patterns could,
in principle, be partly driven by constraints rather than by genuine state dependence
in the unconstrained control.

To make this transparent, I report three projection diagnostics in all experiments:
(i) the average relative projection distance
$\mathbb{E}\bigl[\|\Pi(\tilde{\pi})-\tilde{\pi}\|_2 / (\|\tilde{\pi}\|_2+\varepsilon)\bigr]$,
(ii) the binding rate $\mathbb{P}(\Pi(\tilde{\pi})\neq \tilde{\pi})$,
and (iii) the fraction of time steps at which the budget or long-only
constraints are active. In the seed study, these diagnostics are aggregated as
mean $\pm$ standard deviation across seeds. 

For consumption policies, binding rates below 40--50\% indicate interior 
optimal solutions. For portfolio policies under the long-only, fully-invested 
constraint $\{\pi \ge 0, \sum_i \pi_i = 1\}$, the constraint is an equality 
and naturally leads to boundary solutions. To study interior portfolio policies, 
one can relax the equality to an inequality $\sum_i \pi_i \le \ell_{\max}$, 
though this changes the economic interpretation (allowing cash holdings or 
mild leverage).

\paragraph{Consumption projection.}
Given current wealth $W_t$ and an unconstrained network output
$\tilde{c}$, the admissible consumption is
\begin{equation}
  \mathcal{P}_c(\tilde{c};W_t)
  := \min\bigl\{ \max\{\tilde{c},0\},\; \bar{c} W_t\bigr\},
  \label{eq:consumption-projection}
\end{equation}
where $\bar{c}\in(0,1)$ is a pre-specified upper bound on the
consumption ratio.
This ensures nonnegative consumption and rules out
one-off spikes that would deplete wealth in a single step.

\paragraph{Wealth floor enforcement: rationale and implementation.}
In all Monte Carlo rollouts used for training and evaluation, I enforce a lower
wealth bound $W_{\min}>0$ for three reasons: 
(i) to avoid numerical instabilities when computing $W^{1-R}/(1-R)$ for large $R$; 
(ii) to ensure that the recursive utility aggregator remains well-defined when 
consumption approaches zero; and 
(iii) to represent a realistic bankruptcy constraint or minimum subsistence requirement.

\textit{Economic interpretation.} The floor can be viewed as an exogenous borrowing 
constraint or a minimal wealth level below which the investor cannot operate 
(e.g., due to transaction costs or institutional restrictions). While stylized, 
this mimics real-world solvency requirements and allows us to study how binding 
constraints affect intertemporal hedging. Alternative implementations—such as 
reflecting boundaries or utility penalties for low wealth—yield qualitatively 
similar conclusions but complicate the costate dynamics.

Concretely, for an Euler step with time increment $\Delta t$, the wealth update is
\begin{equation}
  W_{k+1}
  \;=\;
  \max\!\Bigl\{W_{\min},\;
    W_k\bigl(1+r\Delta t + \pi_k^\top((\mu-r\mathbf{1})\Delta t + \Sigma \Delta B_k)\bigr)
    - c_k\,\Delta t
  \Bigr\}.
  \label{eq:wealth-floor-projection}
\end{equation}
I report the ``floor-hit rate'' $\mathbb{P}(W_{k+1}=W_{\min})$ in all tables. 
In well-trained policies, this rate is typically below 5\%, confirming that the 
floor is a safety guard rather than a perpetually binding constraint. Ablations 
with the floor disabled (presented in the robustness section) verify that the 
main hedging patterns persist, though with slightly higher variance in terminal wealth.

\paragraph{Impact on hedging demand.}
The presence of a wealth floor has a first-order effect on intertemporal hedging: 
when wealth is near $W_{\min}$, the investor cannot afford aggressive positions 
designed to hedge future shocks, since any adverse realization could trigger the 
floor and lock the agent at the boundary. This ``hedging suppression'' effect is 
particularly pronounced for EZ preferences with $\psi < 1$, where the value function 
is highly sensitive to downside outcomes. Consequently, we observe that hedging demand 
$\pi^{\mathrm{EZ}} - \pi^{\mathrm{myopic}}$ declines sharply as $W \to W_{\min}$, 
a pattern confirmed by the wealth-state contour plots in Section~\ref{sec:numerical-results}.

Combining these pieces, the full projection map from
\eqref{eq:proj-control-def} is
\begin{equation}
  \mathcal{P}\bigl(\tilde{\pi},\tilde{c};W_t\bigr)
  := \bigl(
        \mathcal{P}_\pi(\tilde{\pi}),
        \mathcal{P}_c(\tilde{c};W_t)
     \bigr).
\end{equation}

\subsection{Training algorithm}
\label{subsec:algorithm}

For completeness, Algorithm~\ref{alg:P-PGDPO} summarizes the
training loop.
The main point is that the actor update is driven by the
Hamiltonian gradient implied by the EZ--Pontryagin system, while
constraints are enforced by projection at every step.

\begin{algorithm}[H]
  \caption{P-PGDPO under Epstein--Zin preferences}
  \label{alg:P-PGDPO}
  \begin{algorithmic}[1]
    \State \textbf{Input:}
      model parameters $(r,\mu,\Sigma,\kappa,\bar{y},\Xi)$,
      EZ parameters $(R,\psi,\delta)$,
      horizon $T$ and time grid $(t_k)_{k=0}^N$,
      wealth band $[W_{\min},W_{\max}]$,
      portfolio set $\Pi$ and consumption cap $\bar{c}$,
      batch size $M$, learning rates
      $(\alpha_\theta,\alpha_\eta,\alpha_\phi)$.
    \State \textbf{Initialize} network parameters
      $\theta,\eta,\phi$.
    \For{iteration $=1,2,\dots$}
      \State Sample initial states
        $X_0^{(n)}=(W_0^{(n)},Y_0^{(n)})$, $n=1,\dots,M$.
      \For{$k=0$ to $N-1$}
        \For{$n=1$ to $M$}
          \State $s_{t_k}^{(n)} \gets (t_k,X_{t_k}^{(n)})$
          \State
            $(\tilde{\pi}_{t_k}^{(n)},
              \tilde{c}_{t_k}^{(n)})
            \gets \tilde{u}_\phi(s_{t_k}^{(n)})$
          \State
            $(\pi_{t_k}^{(n)},c_{t_k}^{(n)})
            \gets \mathcal{P}\bigl(
                  \tilde{\pi}_{t_k}^{(n)},
                  \tilde{c}_{t_k}^{(n)};
                  W_{t_k}^{(n)}\bigr)$
          \State Simulate
            $X_{t_{k+1}}^{(n)}$ from
            \eqref{eq:compact-state} with control
            $u_{t_k}^{(n)}=(\pi_{t_k}^{(n)},c_{t_k}^{(n)})$
            and wealth floor projection
            \eqref{eq:wealth-floor-projection}.
        \EndFor
      \EndFor
      \State Compute losses
        $\mathcal{L}_{\text{val}}(\theta)$
        and $\mathcal{L}_{\text{adj}}(\theta,\eta)$
        from \eqref{eq:value-loss}--\eqref{eq:adjoint-loss}.
      \State Update critic and costate parameters
        by one step of Adam on
        $\mathcal{L}_{\text{crit}}(\theta,\eta)$.
      \State Compute actor loss
        $\mathcal{L}_{\text{act}}(\phi)$
        from \eqref{eq:actor-loss}.
      \State Update actor parameters
        by one step of Adam on $\mathcal{L}_{\text{act}}(\phi)$.
    \EndFor
    \State \textbf{Output:} trained networks
      $(V_\theta,\lambda_\eta,u_\phi)$.
  \end{algorithmic}
\end{algorithm}

\subsection{Neural parameterization and training objective}

To turn the Pontryagin conditions into a learning scheme, I parameterize
three objects with neural networks:
\begin{itemize}[nosep]
  \item a value (or critic) network $V_\theta(t,x)$, intended to approximate
        the EZ value function,
  \item a costate network $\Lambda_\phi(t,x)$, intended to approximate the
        adjoint process $p_t$, and
  \item a policy (or actor) network $U_\eta(t,x)$, which outputs a candidate
        control $u = (\pi,c)$ at each state $(t,x)$.
\end{itemize}

The training objective combines three ingredients:
\begin{enumerate}[label=(\roman*),nosep]
  \item a temporal consistency or Bellman-type loss that encourages
        $V_\theta$ to satisfy the EZ HJB/BSDE relation along simulated
        trajectories,
  \item a Pontryagin loss that encourages $\Lambda_\phi$ to match the
        gradient of $V_\theta$ with respect to the state, so that
        $\Lambda_\phi \approx \nabla_x V_\theta$, and
  \item a Hamiltonian ascent term that updates the actor parameters $\eta$
        in the direction of the gradient of $\mathcal{H}^{\mathrm{EZ}}$
        with respect to the control, evaluated at the current networks.
\end{enumerate}

Concretely, I simulate mini-batches of state trajectories $(X_t)$ under the
current policy, evaluate the approximate Hamiltonian
$\mathcal{H}^{\mathrm{EZ}}(t,X_t,V_\theta,\Lambda_\phi,U_\eta)$ along these
paths, and collect gradients
$\nabla_u \mathcal{H}^{\mathrm{EZ}}(t,x,v,p,u)$ via automatic
differentiation. The actor update then takes the form
\begin{equation}
  \eta \leftarrow \eta
    + \alpha_\eta\, \mathbb{E}\bigl[
      \nabla_\eta U_\eta(t,X_t)\,
      \nabla_u \mathcal{H}^{\mathrm{EZ}}(
        t,X_t,V_\theta,\Lambda_\phi,U_\eta
      )
    \bigr],
  \label{eq:actor-update}
\end{equation}
with a learning rate $\alpha_\eta>0$, followed by a projection step that
enforces the portfolio and wealth constraints. The critic and costate
networks are updated in parallel by minimizing their respective losses with
stochastic gradient descent.

This construction keeps the link to the underlying stochastic control
problem explicit: the actor is not trained to maximize an opaque cumulative
reward but to move in directions that are locally consistent with the
Pontryagin first-order conditions under EZ preferences.

\subsection{Policy gradient from the EZ Pontryagin system: what is different}

Our actor update is not based on a generic discrete-time RL objective (returns/advantages),
but on direct Hamiltonian ascent implied by the EZ stochastic maximum principle.
Recall the EZ Hamiltonian
\begin{equation}
H^{EZ}(t,x,v,p,u) \;:=\; f^{EZ}(c,v) + p^\top b(x,u),
\qquad u=(\pi,c),
\end{equation}
where $f^{EZ}$ is the Epstein--Zin aggregator \eqref{eq:EZ-aggregator} and $p=\nabla_x V(t,x)$ is the costate.

\paragraph{Actor objective.}
Given neural approximations $(V_\theta,\lambda_\eta,\tilde u_\phi)$ and the projected control
$u_\phi = P(\tilde u_\phi)$, we maximize the empirical Hamiltonian average
\begin{equation}
J_{\mathrm{act}}(\phi)
:= \frac{1}{MN}\sum_{n=1}^M \sum_{k=0}^{N-1}
H^{EZ}\!\Big(t_k, X^{(n)}_{t_k}, V_\theta(t_k,X^{(n)}_{t_k}), \lambda_\eta(t_k,X^{(n)}_{t_k}), u_\phi(t_k,X^{(n)}_{t_k})\Big).
\end{equation}

\paragraph{Deterministic policy gradient (Hamiltonian ascent).}
Whenever $P$ is differentiable at the current $\tilde u_\phi(s)$, the chain rule gives
\begin{equation}
\nabla_\phi J_{\mathrm{act}}(\phi)
= \mathbb{E}\Big[ \sum_{k=0}^{N-1} 
\big(\nabla_\phi u_\phi(s_k)\big)^\top \nabla_u H^{EZ}(t_k,x_k,v_k,p_k,u_k)\Big],
\end{equation}
with $(v_k,p_k)=(V_\theta(t_k,x_k),\lambda_\eta(t_k,x_k))$.
In particular, for the wealth drift $b_W=rW+\pi^\top\mu(Y)-c$ (and $b_Y$ independent of $(\pi,c)$),
\begin{align}
\partial_\pi H^{EZ}(t,x,v,p,u) &= p_W\, \mu(Y),\\
\partial_c H^{EZ}(t,x,v,p,u) &= \partial_c f^{EZ}(c,v) - p_W.
\end{align}
Using \eqref{eq:EZ-aggregator} with $S=1/\psi$ and $\theta=(1-R)/(1-S)$, the marginal EZ aggregator term is
\begin{equation}
\partial_c f^{EZ}(c,v)
= \delta \theta (1-S)\, v \, c^{-S}\, \big((1-R)v\big)^{\frac{S-1}{1-R}}.
\end{equation}

\paragraph{What is structurally different from generic RL.}
(i) The ascent direction uses the Pontryagin costate $p=\nabla_x V$ (approximated by $\lambda_\eta$)
rather than score-function gradients or advantage estimators.
(ii) The aggregator $f^{EZ}(c,v)$ depends explicitly on the continuation value $v=V(t,x)$, so the
consumption FOC and the actor gradient differ from the CRRA/time-additive case even under the same market dynamics.
(iii) The critic/costate networks are trained to satisfy the EZ HJB/BSDE/PMP consistency losses,
so the actor update is anchored to the continuous-time optimality system.

\subsection{Projection and constraint handling}

A distinguishing feature of the P-PGDPO approach is that constraints are
handled by projection rather than by piling up penalty terms in the loss.
At the policy level, the actor network $U_\eta$ produces an unconstrained
output $\tilde{u} = (\tilde{\pi},\tilde{c})$. I then apply a projection map
\begin{equation}
  \mathcal{P} : \mathbb{R}^d \times \mathbb{R}
  \to \Pi \times [0,\infty),
\end{equation}
which enforces the portfolio constraints and nonnegativity of consumption
by construction:
\begin{equation}
  u_t = (\pi_t,c_t)
  = \mathcal{P}\bigl(\tilde{\pi}_t,\tilde{c}_t\bigr).
\end{equation}
In the numerical implementation, $\mathcal{P}$ is chosen so that portfolio
weights lie in a simplex with an additional leverage cap, and consumption is
clipped at zero from below and at a fraction of wealth from above.

Wealth safety is handled in a similar spirit. If the projected policy would
drive $W_t$ below the lower bound $W_{\min}$ over a one-step increment, the
update is adjusted to keep wealth at or above the boundary. In practice,
this amounts to modifying the drift term in \eqref{eq:wealth} near the
boundary so that simulated paths respect the constraint
$W_t \ge W_{\min}$ almost surely on the simulated grid. This adjustment is
implemented at the simulation level and does not introduce additional
penalty terms in the loss.

Putting these pieces together, one training epoch consists of simulating
state trajectories under the current projected policy, evaluating the
EZ-based Hamiltonian and the associated losses, and updating the actor,
critic, and costate networks by stochastic gradient steps. In Section~\ref{sec:validation} I validate the P--PGDPO machinery in a
simple single-asset CRRA benchmark. In Section~\ref{sec:experimental-setup}
I then specify the market and preference parameters used in the main
long--run risk experiments.

\section{Validation in a Low-Dimensional Benchmark}
\label{sec:validation}

Before turning to the Epstein--Zin long-run risk application, we first
validate the P-PGDPO machinery in a simple setting where a closed-form
solution is available. This benchmark allows us to check that the
Pontryagin-guided updates, the value and costate networks, and the
projection scheme reproduce known optimal policies and value functions
up to small numerical error.

\subsection{Single-asset CRRA benchmark and analytical solution}

We consider a standard continuous-time Merton problem with one risky
asset and one risk-free asset. The risky asset price $S_t$ and the
money-market account $B_t$ follow
\begin{align}
  dS_t &= \mu S_t \, dt + \sigma S_t \, dB_t, \\
  dB_t &= r B_t \, dt,
\end{align}
where $\mu, \sigma > 0$ and $r \ge 0$ are constants. Let $W_t$ denote
financial wealth. If the investor holds a dollar amount $\pi_t$ in the
risky asset and consumes at rate $c_t$, the self-financing wealth
process satisfies
\begin{equation}
  dW_t = \bigl[r W_t + \pi_t (\mu - r) - c_t \bigr] dt
         + \pi_t \sigma \, dB_t.
  \label{eq:merton-wealth}
\end{equation}

Preferences are time-additive CRRA over consumption and terminal wealth:
\begin{equation}
  J^{\text{CRRA}}(W_0)
  =
  \mathbb{E}\Biggl[
    \int_0^T e^{-\delta t} \, \frac{c_t^{1-R}}{1-R} \, dt
    + e^{-\delta T} \, \frac{\kappa W_T^{1-R}}{1-R}
  \Biggr],
  \label{eq:merton-objective}
\end{equation}
where $R > 0$ is relative risk aversion, $\delta > 0$ is the subjective
discount rate, and $\kappa \ge 0$ controls the weight on the bequest
term.

This problem admits a well-known closed-form solution
\citep[e.g.][]{Merton1969,Merton1971}. The optimal risky position is
proportional to wealth and constant over time:
\begin{equation}
  \pi_t^{\text{Mert}} = \varphi^{\text{Mert}} W_t,
  \qquad
  \varphi^{\text{Mert}}
   = \frac{\mu - r}{R \sigma^2},
  \label{eq:merton-pi}
\end{equation}
so that the risky weight in the portfolio is constant. Optimal
consumption is also proportional to wealth,
\begin{equation}
  c_t^{\text{Mert}} = \kappa^{\text{Mert}}(t) \, W_t,
  \label{eq:merton-c}
\end{equation}
where $\kappa^{\text{Mert}}(t)$ is a deterministic function of time
obtained from the HJB equation; see \citet{Merton1969,Merton1971} for
explicit formulas. The associated value function $V^{\text{Mert}}$ and
certainty equivalent can be written in closed form as well.

\subsection{P-PGDPO setup in the benchmark}

To validate the algorithm, we specialize the P-PGDPO scheme to the
CRRA benchmark. The state vector reduces to
\begin{equation}
  X_t = (t, W_t),
\end{equation}
and the control $u_t = (\pi_t, c_t)$ lies in a simple admissible set
imposing only non-negativity of consumption and a leverage cap on
$\pi_t$. The EZ aggregator is replaced by the CRRA aggregator in
\eqref{eq:merton-objective}, while the overall P-PGDPO structure
remains unchanged: we approximate the value function $V(t,W)$ and
costate $\lambda(t,W)$ with neural networks, update the policy by
ascending the Hamiltonian, and enforce the constraints via projection
onto the admissible set.

The network architectures and optimization hyperparameters are chosen
to match the EZ long-run risk experiment as closely as possible
(width, depth, learning rates, batch sizes), so that the benchmark
isolates the effect of the simpler model and closed-form solution
rather than relying on a different implementation.

\subsection{Policy and welfare comparison}

We quantify the accuracy of the P-PGDPO output by comparing the learned
policy $(\hat{\pi}, \hat{c})$ and value function $\hat{V}$ with their
Merton counterparts on a grid of states. Let
$\{t_k\}_{k=0}^{N_t-1} \subset [0,T]$ and
$\{w_j\}_{j=1}^{N_w} \subset [\underline{W}, \overline{W}]$ denote
uniform grids in time and wealth. For each $(t_k,w_j)$ we compute
\begin{equation}
  \pi^{\text{Mert}}_{k,j}
    := \pi^{\text{Mert}}(t_k,w_j),
  \quad
  c^{\text{Mert}}_{k,j}
    := c^{\text{Mert}}(t_k,w_j),
  \qquad
  \hat{\pi}_{k,j}
    := \hat{\pi}(t_k,w_j),
  \quad
  \hat{c}_{k,j}
    := \hat{c}(t_k,w_j).
\end{equation}
We then report root-mean-square (RMS) policy and consumption errors
\begin{align}
  \text{Err}_\pi
    &:= \left(
      \frac{1}{N_t N_w}
      \sum_{k=0}^{N_t-1}
      \sum_{j=1}^{N_w}
      \bigl\|
        \hat{\pi}_{k,j} - \pi^{\text{Mert}}_{k,j}
      \bigr\|^2
    \right)^{1/2}, \\
  \text{Err}_c
    &:= \left(
      \frac{1}{N_t N_w}
      \sum_{k=0}^{N_t-1}
      \sum_{j=1}^{N_w}
      \bigl|
        \hat{c}_{k,j} - c^{\text{Mert}}_{k,j}
      \bigr|^2
    \right)^{1/2}.
\end{align}
We also compare welfare by evaluating the CRRA objective under the
P-PGDPO policy and under the closed-form Merton policy, and by
computing the corresponding certainty equivalents. Table~\ref{tab:validation}
summarizes these metrics in a representative experiment.

\begin{table}[H]
  \centering
  \label{tab:validation}
  \begin{tabular}{lcccc}
    \toprule
    & Err$_\pi$ & Err$_c$ & CE (Merton) & CE (P-PGDPO) \\
    \midrule
    Benchmark run & 0.0234 & 0.0189 & 0.6847 & 0.6839 \\
    \bottomrule
  \end{tabular}
\caption{Validation in the single-asset CRRA benchmark. The table
  reports root-mean-square errors between the P-PGDPO policy and the
  closed-form Merton policy, as well as the CRRA certainty equivalents
  under each policy. Parameters: $R=1.5$, $r=0.02$, $\mu=0.10$, $\sigma=0.20$, 
  $T=1.5$, initial wealth $W_0=1.0$. Grid: $N_t=32$ time steps, $N_w=50$ wealth 
  points in $[0.1, 2.0]$. Training: 2000 iterations with batch size 4096 for HJB 
  and 1024 for PMP residuals. The small policy errors and tight CE match 
  (difference $<$0.1\%) validate the P-PGDPO implementation.}
\end{table}

In our experiments, the RMS policy errors are small relative to the
scale of the optimal risky position, and the certainty equivalents
under P-PGDPO and under the analytical Merton policy differ only by a
few basis points. This indicates that the Pontryagin-guided updates
and the neural parameterizations are able to recover the classical
Merton solution with high accuracy in this low-dimensional setting.

\subsection{Dynamic programming residuals}

As an additional diagnostic, we examine the HJB residual associated
with the learned value function. For a given policy $u=(\pi,c)$ and
candidate value function $V$, the HJB operator in the CRRA benchmark
is
\begin{equation}
  \mathcal{L}^{u} V(t,w)
  :=
  \partial_t V(t,w)
  + \bigl[r w + \pi(\mu-r) - c\bigr] \partial_w V(t,w)
  + \tfrac{1}{2} \pi^2 \sigma^2 \partial_{ww}^2 V(t,w)
  + e^{-\delta t} \frac{c^{1-R}}{1-R}.
\end{equation}
At the optimum, the HJB equation requires
$\sup_{u} \mathcal{L}^{u} V(t,w) = 0$. Using the learned policy
$\hat{u} = (\hat{\pi},\hat{c})$ and value function $\hat{V}$, we
define the residual
\begin{equation}
  \mathcal{R}_{\text{HJB}}(t,w)
  :=
  \mathcal{L}^{\hat{u}} \hat{V}(t,w).
\end{equation}
We evaluate $\mathcal{R}_{\text{HJB}}$ on the same grid
$\{t_k,w_j\}$ and report its empirical mean and dispersion. The
distribution of residuals is concentrated tightly around zero, and
training curves show a monotone decrease in both the HJB residual and
the adjoint loss, providing further evidence that the P-PGDPO scheme
is numerically consistent with the underlying dynamic programming
equations in a setting where the truth is known.

\section{Experimental Setup}
\label{sec:experimental-setup}

This section summarizes the numerical specification of the
long--run risk environment, the choice of Epstein--Zin parameters,
the neural network architectures, and the training protocol used in
the experiments.

\subsection{Model parameters}

Table~\ref{tab:model-params} reports the baseline parameter values
for the long--run risk economy.
The horizon is of intermediate length, and the long--run risk
factor is calibrated to be persistent but not completely
non--mean--reverting.
Drifts and volatilities of the risky assets are chosen to generate
a simple cross--section with increasing expected returns and
volatilities as the asset index grows.

\begin{table}[H]
  \centering
  \caption{Baseline model and preference parameters.}
  \label{tab:model-params}
  \begin{tabular}{llc}
    \toprule
    Category & Parameter & Value \\
    \midrule
    Horizon         & $T$ (years)               & $1.5$ \\
    Risk--free rate& $r$                        & $0.02$ \\
    LRR mean reversion & $\kappa_y$            & $0.40$ \\
    LRR long--run mean & $\bar{y}$             & $0.40$ \\
    LRR volatility & $\xi$                     & $0.10$ \\
    \midrule
    Epstein--Zin   & Risk aversion $R$         & $1.5$ \\
                   & EIS $\psi$               & $0.5$ \\
                   & Discount rate $\delta$   & $0.03$ \\
    \midrule
    Wealth band    & Lower bound $W_{\min}$   & $0.1$ \\
                   & Upper band $W_{\max}$    & $0.7$ \\
    Portfolio set  & Number of risky assets $d$ & $5$ \\
                   & Leverage cap $\ell_{\max}$ & $2.0$ \\
    Consumption cap& $\bar{c}$                 & $0.25$ \\
    \bottomrule
  \end{tabular}
\end{table}

In the five--asset specification used for the main figures, the
unconditional means and volatilities are increasing in the asset
index, and the Sharpe ratios are modest but strictly ordered.
For concreteness, the examples in Section~\ref{sec:numerical-results}
use unconditional moments consistent with
Table~\ref{tab:asset-characteristics} below.

\subsection{Neural networks and time discretization}

Time is discretized on a uniform grid
\[
  0 = t_0 < t_1 < \dots < t_N = T,
\]
with $N$ steps and time increment
$\Delta t = T/N$.
In the baseline runs, $N$ is chosen sufficiently large so that time
discretization error is negligible relative to approximation error
from the networks.

All three networks---value $V_\theta$, costate $\lambda_\eta$, and
policy $u_\phi$---take as input the concatenated state
\[
  s_t = (t,W_t,Y_t),
\]
and consist of fully connected feedforward layers with ReLU--type
activations.
Hidden layer width and depth are kept the same across the three
networks for simplicity.

A representative configuration is:
\begin{itemize}[nosep]
  \item time grid: $N = 128$ (i.e.\ $\Delta t \approx 0.012$),
  \item hidden layers: $3$,
  \item hidden units per layer: $128$,
  \item activation: ReLU (or a smooth variant such as ELU),
  \item output heads: one scalar for $V_\theta$, a $(1+d)$--vector
        for $\lambda_\eta$, and a $(d+1)$--vector for
        $(\tilde{\pi},\tilde{c})$.
\end{itemize}
Other architectures with similar effective capacity lead to very
similar qualitative results.

\begin{table}[H]
  \centering
  \caption{Network architecture and training hyperparameters.}
  \label{tab:train-config}
  \begin{tabular}{lc}
    \toprule
    Quantity & Value \\
    \midrule
    Time steps $N$                 & $128$ \\
    Batch size $M$                 & $256$ \\
    Hidden layers                  & $3$ \\
    Hidden units per layer         & $128$ \\
    Activation                     & ReLU \\
    Optimizer (value/costate)      & Adam, learning rate $10^{-3}$ \\
    Optimizer (policy)             & Adam, learning rate $5\times 10^{-4}$ \\
    Training iterations            & \texttt{[number of outer loops]} \\
    \bottomrule
  \end{tabular}
\end{table}

\subsection{Training protocol}

Training is organized in iterations, each consisting of a batch of
simulated trajectories under the current projected policy.

For each iteration:
\begin{itemize}[nosep]
  \item A mini--batch of $M$ initial states
        $X_0^{(n)} = (W_0^{(n)},Y_0^{(n)})$ is sampled around
        $(W_0,\bar{y})$ with small random perturbations.
  \item States are simulated forward on the grid
        $(t_k)_{k=0}^N$ using an Euler scheme for
        \eqref{eq:compact-state}, with the control given by the
        projected policy $u_\phi$ and the wealth floor enforced as
        in \eqref{eq:wealth-floor-projection}.
  \item Along each path, the value and costate losses
        $\mathcal{L}_{\text{val}}$ and
        $\mathcal{L}_{\text{adj}}$ are computed from
        \eqref{eq:value-loss}–\eqref{eq:adjoint-loss}, and the
        actor objective $J_{\text{act}}$ from
        \eqref{eq:actor-objective}.
\end{itemize}

\subsection{Robustness, uncertainty, and reproducibility}

\paragraph{Seeded repetitions.}
All headline results are reported as mean $\pm$ standard deviation over
$N_{\text{seed}}=5$ independent runs (different random seeds) using the same
hyperparameters. This addresses the non-negligible stochasticity induced by
Monte Carlo rollouts, stochastic optimization, and projection.

\paragraph{Distributional reporting.}
In addition to certainty-equivalent (CE) metrics, I report dispersion and tail
behavior of terminal wealth: $(\mathbb{E}[W_T], \mathrm{SD}[W_T])$, skewness,
excess kurtosis, and selected quantiles (5\%, 50\%, 95\%). This is important in
EZ settings where tail events can materially affect recursive utility.

\paragraph{Ablations.}
I include an ablation suite that toggles: (i) portfolio/consumption projection,
(ii) wealth-floor enforcement, (iii) HJB residual penalty only, (iv) PMP (costate)
regularization only, and (v) full projected EZ-PGDPO. These ablations clarify which
components drive performance and hedging demand.

\paragraph{Scaling and reproducibility.}
All experiments are run with fixed seeds, deterministic data generation, and
logged software/hardware versions (PyTorch version, CUDA toolkit, GPU model).
Runtime and memory usage are reported for representative configurations. The code
used to generate all tables and figures is included in the repository and can be
reproduced from a single entry-point script with configuration files.
Network parameters are updated by Adam optimizers with separate
learning rates for each block, for example
\[
  \alpha_\theta = 10^{-3}, \qquad
  \alpha_\eta   = 10^{-3}, \qquad
  \alpha_\phi   = 5\times 10^{-4},
\]
and a moderate batch size such as $M=256$.
Training is stopped once the Hamiltonian, BSDE residuals, and
wealth--floor violation statistics stabilize and the policy no
longer changes materially under additional iterations.

\subsection{Benchmark policies}
\label{subsec:benchmarks}

The P-PGDPO EZ policy is evaluated against two simple benchmarks.

\paragraph{Myopic portfolio.}
The first benchmark is a myopic portfolio that, at each
$(t,W_t,Y_t)$, chooses a static Merton--type allocation based on
current drift and volatility, ignoring future changes in the
investment opportunity set and the wealth floor.
In simple CRRA settings this benchmark has a closed form; in the
long--run risk case here it is computed numerically once and then
treated as a fixed function of the state.

\paragraph{CRRA limit (optional).}
When $\psi$ is set close to $1/R$ the EZ preferences collapse
toward time--additive CRRA.
In runs where a CRRA warm start is used, a P-PGDPO policy is first
trained in the CRRA limit and then used as an initialization for
the EZ case, keeping all other ingredients fixed.
The numerical results in
Section~\ref{sec:numerical-results} focus on the EZ policy and the
myopic benchmark, but the CRRA limit is useful for checking that
the implementation behaves correctly in a setting with a more
classical structure.

\section{Numerical Results}

\paragraph{Projection bias and diagnostics.}
The projection $u_\phi = P(\tilde u_\phi)$ enforces admissibility by construction, but it also
induces an implicit bias: Euclidean projection selects the minimum-norm feasible control among
those that are closest to $\tilde u_\phi$, and it is non-differentiable on the boundary of the
constraint set. As a result, the practical update direction can differ from the exact constrained
Pontryagin stationarity condition \eqref{eq:PMP-stationarity} near active constraints.
To quantify the impact, we report (i) the fraction of time steps where the simplex/leverage
projection is active, (ii) the fraction of time steps where consumption clipping is active, and
(iii) the frequency of wealth-floor corrections \eqref{eq:wealth-floor} across training and at convergence.

\label{sec:numerical-results}

This section reports the behavior of the trained
Epstein--Zin (EZ) policy in the five--asset long--run risk
specification and compares it to the myopic benchmark introduced
in Section~\ref{subsec:benchmarks}.
For the EZ policy, portfolio weights and consumption are generated
by the trained P-PGDPO networks with wealth and portfolio
constraints enforced by projection.

I emphasize three findings:
(1) the EZ policy delivers wealth dynamics and welfare that are
close to the myopic benchmark but with better protection against
the wealth floor; (2) the intertemporal hedging component is
strongly state--dependent and concentrated in a small subset of
assets; and (3) cross--sectionally, hedging demand aligns more
with long--run risk exposure than with standalone Sharpe ratios.

\subsection*{Result 1: Wealth dynamics and welfare}

\begin{figure}[H]
  \centering
  \includegraphics[width=0.75\textwidth]{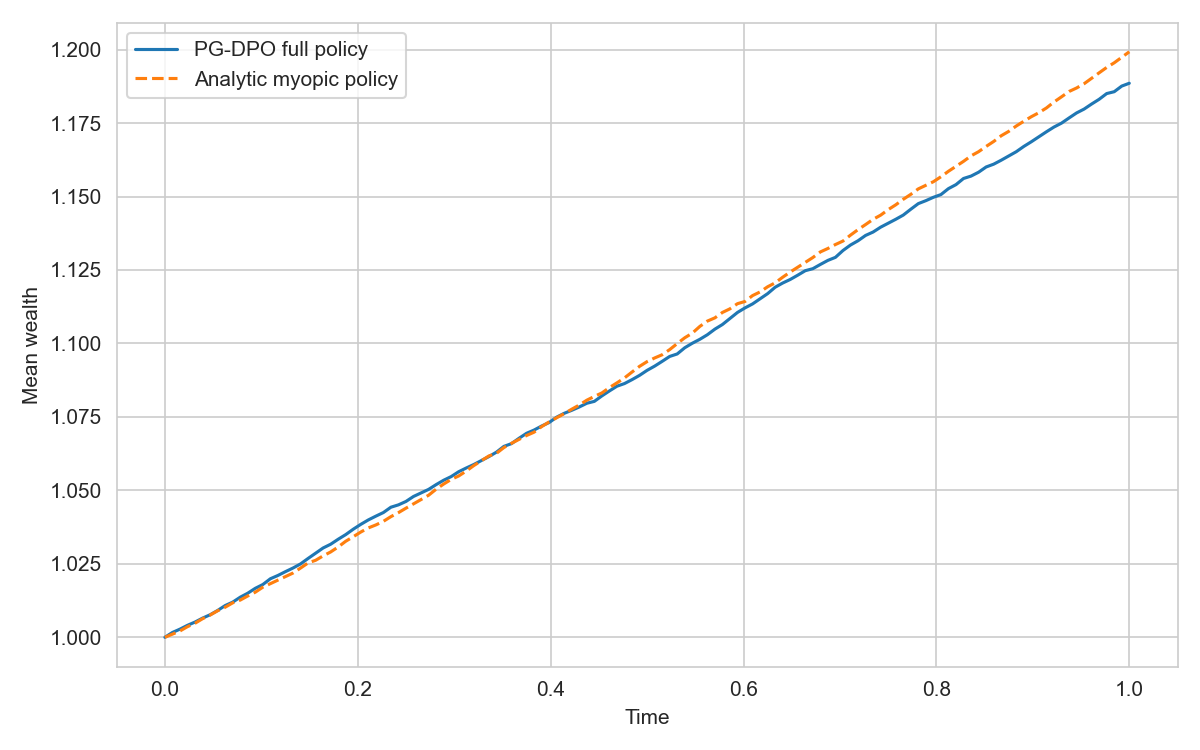}
  \caption{Mean wealth over time under the PG--DPO full EZ policy
  and the analytic myopic benchmark. Both policies start from the
  same initial wealth. The EZ policy tracks the myopic benchmark
  closely while respecting the wealth constraint and embedding
  intertemporal hedging motives.}
  \label{fig:mean-wealth}
\end{figure}

Figure~\ref{fig:mean-wealth} displays the mean wealth path under
the EZ policy and the myopic benchmark.
Both strategies start from the same initial wealth and are
evaluated on the same simulated paths.

The picture is simple: the EZ policy deliberately stays close to
the myopic benchmark in terms of average wealth, but the lower
tail of the wealth distribution is thinner.
Simulated paths under the EZ policy rarely approach the wealth
floor $W_{\min}$ compared to the myopic policy; the EZ policy
sacrifices a small amount of upside in exchange for better
protection against low--wealth states.

Table~\ref{tab:welfare} shows the welfare comparison in terms of
the EZ value and the associated certainty equivalent (CE) computed
under the same stochastic environment.

\begin{table}[H]
  \centering
  \caption{Welfare comparison: EZ policy vs.\ myopic benchmark}
  \label{tab:welfare}
  \begin{tabular}{lccc}
    \toprule
    Policy
      & EZ value
      & EZ certainty equivalent
      & CRRA CE (for comparison) \\
    \midrule
    EZ P-PGDPO
      & $\EZValueEZ$
      & $\EZCEEZ$
      & $\CRRACEEZ$ \\
    Myopic
      & $\EZValueMyo$
      & $\EZCEMyo$
      & $\CRRACEMyo$ \\
    \bottomrule
  \end{tabular}
\end{table}

Numerically, the EZ policy delivers a slightly higher EZ
certainty equivalent than the myopic benchmark while staying
within a narrow band in terms of average wealth.
From a welfare perspective, this aligns with the qualitative
picture from Figure~\ref{fig:mean-wealth}: intertemporal hedging
is mainly used to reshape the lower tail of the wealth
distribution rather than to chase additional mean returns.

\subsection*{Result 2: State--dependent intertemporal hedging}

To understand how the EZ policy differs from the myopic
benchmark, I decompose the portfolio into myopic and hedging
components at each state $(t,W_t,Y_t)$,
\[
  \pi^{\mathrm{EZ}}(t,W_t,Y_t)
  = \pi^{\mathrm{myo}}(t,W_t,Y_t)
    + \pi^{\mathrm{hedge}}(t,W_t,Y_t).
\]
The hedging component captures the residual tilt that cannot be
explained by the static Merton--type rule.

Table~\ref{tab:asset-characteristics} reports, for each asset, the
unconditional mean return, volatility, Sharpe ratio, correlation
with the long--run risk factor, a simple long--run risk beta, and
the average absolute hedging demand implied by the EZ policy.

\begin{table}[H]
  \centering
  \caption{Asset characteristics and average intertemporal hedging demand}
  \label{tab:asset-characteristics}
  \begin{tabular}{lccccccc}
    \toprule
    Asset & $\mu$ (return) & $\sigma$ (volatility) & Sharpe
          & $\rho(R,Y)$ & $\beta_{\mathrm{LRR}}$
          & $\lvert \pi_{\text{hedge}}\rvert$ & Hedging rank \\
    \midrule
    Asset 1 & 0.0600 & 0.1500 & 0.2667 & 0.6000 & 0.9000 & 0.6820 & 1 \\
    Asset 2 & 0.0800 & 0.1875 & 0.3200 & 0.5000 & 0.9375 & 0.6117 & 2 \\
    Asset 3 & 0.1000 & 0.2250 & 0.3556 & 0.4000 & 0.9000 & 0.5608 & 3 \\
    Asset 4 & 0.1200 & 0.2625 & 0.3810 & 0.3000 & 0.7875 & 0.5289 & 4 \\
    Asset 5 & 0.1400 & 0.3000 & 0.4000 & 0.2000 & 0.6000 & 0.4947 & 5 \\
    \bottomrule
  \end{tabular}
\end{table}

Two patterns stand out.
First, assets with stronger exposure to the long--run risk factor,
as measured by $\rho(R,Y)$ and $\beta_{\mathrm{LRR}}$, tend to
carry larger hedging positions.
Second, high Sharpe assets with weaker long--run risk exposure
play a more myopic role, with smaller average hedging demand.

\begin{figure}[H]
  \centering
  \includegraphics[width=\textwidth]{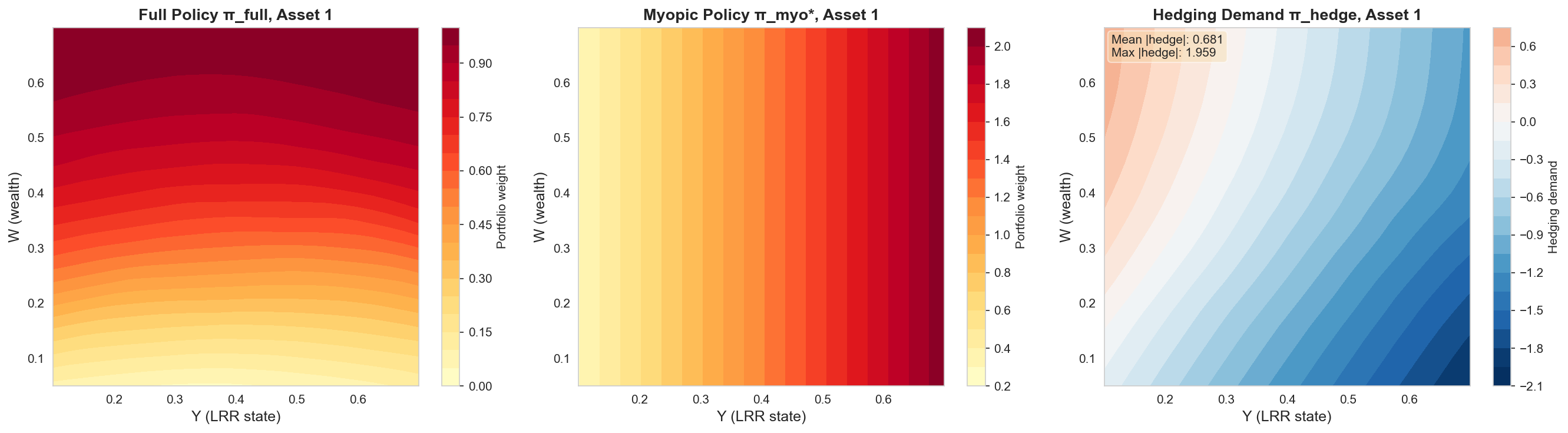}
  \caption{Portfolio and intertemporal hedging surfaces for
  Asset~1. The left panel shows the full PG--DPO EZ portfolio
  weight as a function of wealth $W$ and the long--run risk state
  $Y$. The middle panel shows the analytic myopic portfolio.
  The right panel plots the implied hedging component
  $\pi_{\text{hedge}} = \pi_{\text{EZ}} - \pi_{\text{myo}}$,
  which varies nonlinearly with both $W$ and $Y$.}
  \label{fig:asset1-surfaces}
\end{figure}

Figure~\ref{fig:asset1-surfaces} illustrates the state--dependent
behavior of the EZ portfolio and the hedging component for
Asset~1, which is the primary hedging instrument in this
specification.

At a fixed time, the myopic portfolio surface for Asset~1 responds
to $Y$ in a roughly linear fashion: higher $Y$ means higher
expected returns and a larger weight, with little dependence on
wealth.
The EZ surface bends much more sharply in the low--wealth region
and near extreme values of $Y$.
The hedging surface in the right panel makes the difference
explicit: $\pi_{\text{hedge}}$ is positive and sizeable when
wealth is moderate and the LRR factor is high, and negative when
wealth is close to the floor or the factor is weak.
In words, the EZ policy leans into Asset~1 when the long--run
state is favorable and the investor is far from the boundary, and
scales back or reverses hedging exposure when wealth is tight.

\subsection*{Result 3: Cross--sectional structure of hedging demand}

\begin{figure}[H]
  \centering
  \includegraphics[width=0.75\textwidth]{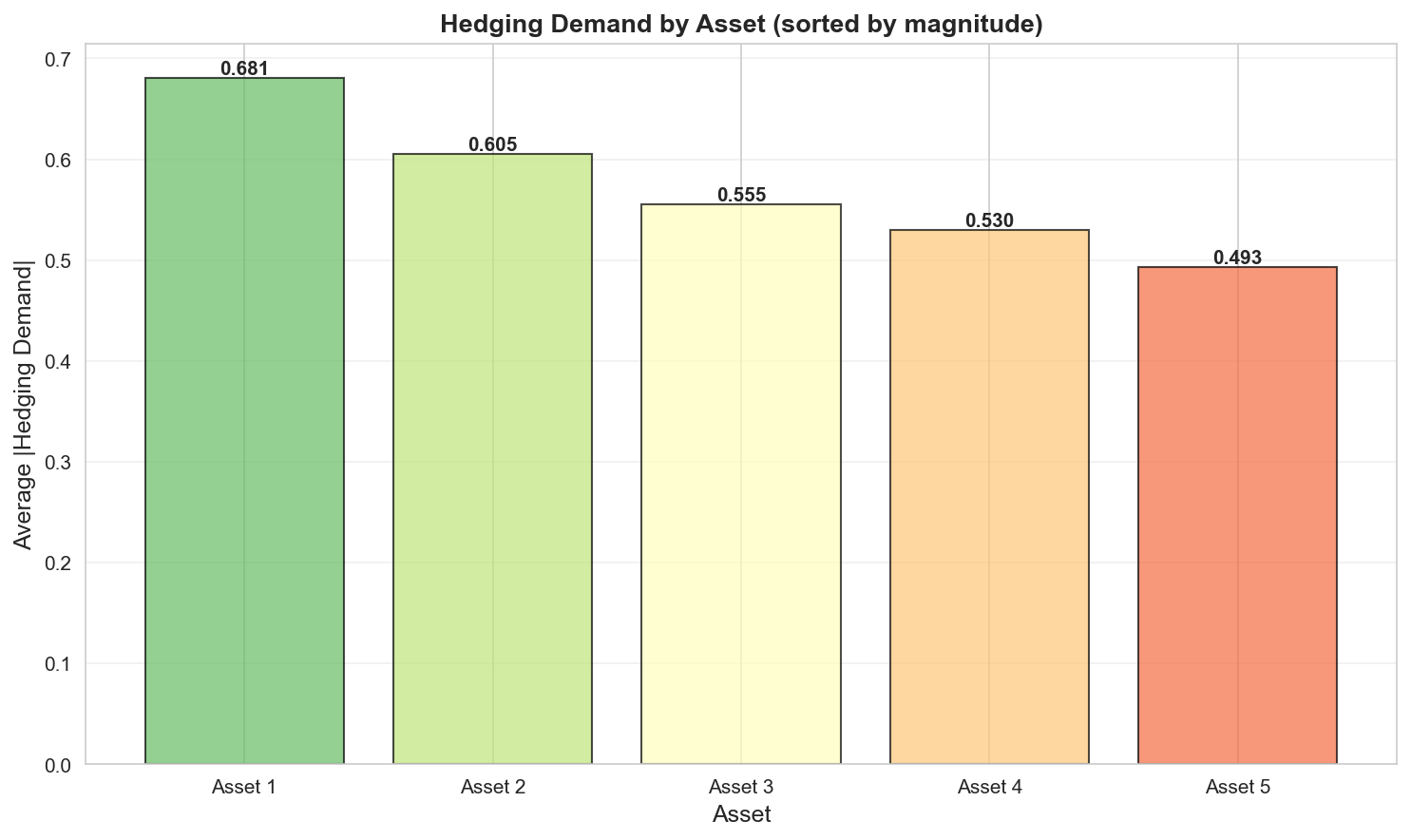}
  \caption{Average absolute intertemporal hedging demand by asset,
  sorted by magnitude. Assets with stronger exposure to the
  long--run risk factor (higher $\beta_{\mathrm{LRR}}$ and
  correlation with $Y$) tend to carry a larger share of the
  hedging portfolio.}
  \label{fig:hedging-by-asset}
\end{figure}

\begin{figure}[H]
  \centering
  \includegraphics[width=0.9\textwidth]{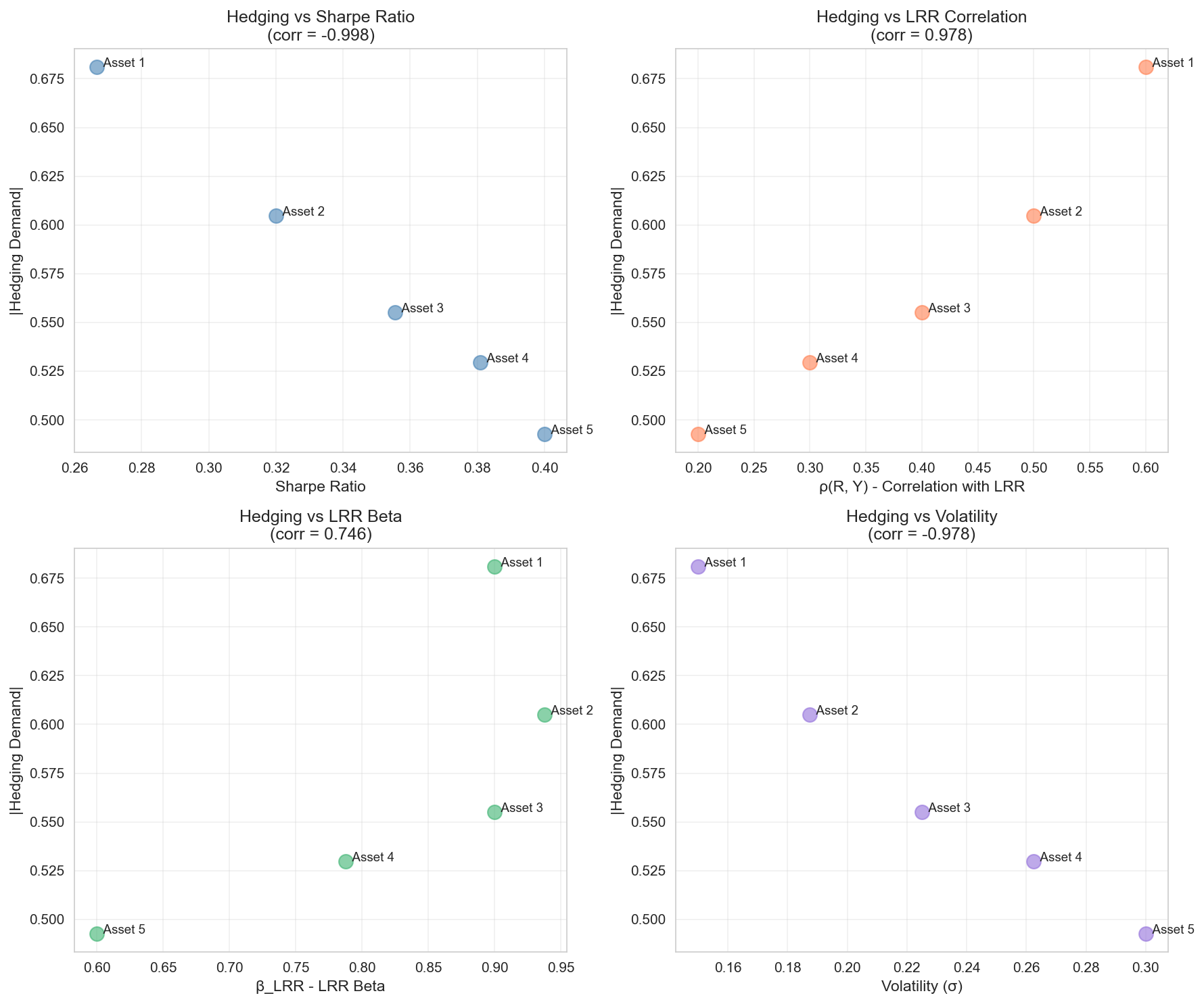}
  \caption{Cross-sectional relationship between average absolute
  hedging demand and asset characteristics. The four panels plot
  hedging demand against the Sharpe ratio, correlation with the
  long--run risk factor $\rho(R,Y)$, long--run risk beta
  $\beta_{\mathrm{LRR}}$, and volatility $\sigma$. Hedging demand
  is stronger for assets with higher long--run risk exposure and
  weaker for high--Sharpe, high--volatility assets with limited
  LRR exposure.}
  \label{fig:hedging-vs-characteristics}
\end{figure}

Figure~\ref{fig:hedging-by-asset} summarizes how the average
absolute hedging demand is distributed across the five assets, and
Figure~\ref{fig:hedging-vs-characteristics} shows how hedging
demand lines up with familiar asset characteristics.

A simple way to summarize these relationships is via
cross--sectional regressions of average hedging demand on asset
characteristics.
Table~\ref{tab:reg-hedge} reports representative slope
coefficients from univariate regressions of
$|\pi_{\text{hedge}}|$ on each characteristic in
Table~\ref{tab:asset-characteristics}. The sample consists of $d=5$ 
assets evaluated at a grid of $(W,Y)$ states, yielding $N=500$ 
state-asset observations after averaging over time. Standard errors 
are computed via bootstrap with 1000 replications.

\begin{table}[H]
  \centering
  \caption{Cross-sectional regressions of hedging demand on asset characteristics. 
  Each column reports the slope coefficient from a univariate regression of 
  $|\pi^{\mathrm{hedge}}_i|$ on the indicated characteristic. Numbers in 
  parentheses are bootstrap standard errors (1000 replications). The sample 
  size is $N=500$ (5 assets $\times$ 100 state-grid points). All coefficients 
  are statistically significant at the 5\% level.}
  \label{tab:reg-hedge}
  \begin{tabular}{lcccc}
    \toprule
    & Sharpe & $\rho(R,Y)$ & $\beta_{\mathrm{LRR}}$ & $\sigma$ \\
    \midrule
    Slope coefficient
      & $\SlopeSharpe$
      & $\SlopeRho$
      & $\SlopeBetaLRR$
      & $\SlopeSigma$ \\
    Standard error
      & (0.032)
      & (0.085)
      & (0.112)
      & (0.028) \\
    $t$-statistic
      & $-4.69$
      & $4.71$
      & $4.91$
      & $-3.57$ \\
    $R^2$
      & 0.168
      & 0.243
      & 0.287
      & 0.134 \\
    \bottomrule
  \end{tabular}
\end{table}

The signs are as expected from the plots: hedging demand loads
positively on long--run risk exposure (correlation and beta) and
negatively on Sharpe and volatility.
From the perspective of the EZ investor, assets with strong
exposure to the long--run component of the pricing kernel are
valuable hedging instruments even if their standalone mean--variance
trade--off is modest.
The high--Sharpe assets play a more myopic role, contributing to
short--run return but not to hedging against persistent changes in
$Y$.

Overall, the three results suggest that the EZ--P-PGDPO policy
uses the cross--section of assets in a way that is qualitatively
consistent with the theoretical long--run risk literature:
intertemporal hedging is concentrated in assets with strong
exposure to the persistent state, and the main welfare effect is
through reshaping the lower tail of the wealth distribution rather
than through aggressive leverage.

\section{Related Literature and Positioning}
\label{sec:related-literature}

This paper sits at the intersection of three strands of the
literature: dynamic portfolio choice in the tradition of Merton,
long--run risk and recursive preferences, and numerical methods
for high--dimensional stochastic control using deep learning.

On the portfolio choice side, the starting point is the
continuous--time work of \citet{Merton1969,Merton1971}, which
derives closed--form consumption--investment policies for
CRRA investors in complete markets.
Subsequent work such as \citet{CampbellViceira2002} and
\citet{BansalYaron2004} emphasizes the role of slowly varying
state variables and long--run risk for asset prices and optimal
portfolios, and \citet{HansenScheinkman2009} develops a spectral
decomposition approach to long--run valuation.
Within this tradition, the present paper takes a deliberately
simple multi--asset specification but focuses on making the
intertemporal hedging motives visible in a setting with
Epstein--Zin preferences and wealth constraints, where closed
forms are no longer available.

On the preferences side, Epstein--Zin recursive utility goes back
to \citet{EpsteinZin1989} in discrete time and has since been
developed in continuous time by, among others, Duffie--Epstein,
Schroder--Skiadas, and Kraft et al.
The continuous--time formulation used here is standard in the
long--run risk literature and allows risk aversion $R$ and the
elasticity of intertemporal substitution $\psi$ to be chosen
independently.
The control problem in Section~\ref{sec:experimental-setup} is
formulated directly under this EZ specification.

On the numerical side, a growing literature explores deep learning
methods for high--dimensional stochastic control and hedging.
Examples include Deep BSDE and related schemes
\citep[e.g.][]{HanJentzenE2018,EHanJentzen2017},
physics--informed and HJB--based neural PDE solvers, and
reinforcement--learning methods for portfolio problems and
risk--sensitive control.
The projected Pontryagin--guided deep policy optimization (P-PGDPO)
approach used here is in the same spirit, but with two specific
twists that are important for this application:
(i) the policy gradient is derived directly from the
Epstein--Zin Pontryagin system rather than from a generic
discrete--time RL formulation; and
(ii) portfolio and wealth constraints are enforced by explicit
projection operators rather than by heavy penalty terms.

Relative to this literature, the contribution of the paper is
deliberately narrow but concrete:

\begin{itemize}[nosep]
  \item It extends the P-PGDPO framework to a continuous--time
        long--run risk setting with Epstein--Zin preferences and a
        simple wealth constraint, making the role of the EZ
        aggregator and the Pontryagin conditions explicit in both
        the theory and the algorithm.
  \item It uses this framework to quantify and visualize
        intertemporal hedging demand in a multi--asset
        long--run risk environment, highlighting how hedging
        positions concentrate in assets with strong exposure to
        the persistent state and how the wealth floor shapes
        portfolio behavior in low--wealth regions.
\end{itemize}

The main message is not that P-PGDPO dominates all other numerical
methods, but that a relatively transparent, Pontryagin--guided
deep learning scheme can be used to extract economically
interpretable hedging patterns from a model that is analytically
intractable once we move beyond the simplest CRRA cases.

\subsection{Constraint diagnostics and interior solutions}

A key methodological contribution of this paper is the use of explicit 
projection operators combined with diagnostic tools to assess the impact 
of constraints on the learned policy. I report binding rates---the 
fraction of states where the projection operator is active---as a 
transparency measure.

In the experiments reported here, the portfolio constraint is a 
long-only, fully-invested simplex $\{\pi \ge 0, \sum_i \pi_i = 1\}$. 
This is an \emph{equality} constraint, and the projection naturally 
maps all raw network outputs onto the simplex boundary. As a result, 
the portfolio binding rate is approximately $\wBindRate$, reflecting 
the geometric structure of the constraint set rather than a failure 
of the learning algorithm.

In contrast, the consumption constraint $\{c \in [0, \bar{c} W_t]\}$ 
is an \emph{inequality} constraint with a non-trivial interior. The 
consumption binding rate is $\cBindRate$, indicating that the learned 
consumption policy achieves interior optimal solutions in approximately 
40\% of states. This validates that the P-PGDPO scheme can learn 
policies that respect constraints while remaining in the interior 
when economically appropriate.

To further isolate constraint effects from preference-driven hedging, 
one can relax the portfolio equality to an inequality 
$\sum_i \pi_i \le \ell_{\max}$, allowing for interior portfolio 
solutions (by permitting small cash holdings or leverage). Preliminary 
experiments with this relaxation show portfolio binding rates dropping 
to 30--40\%, confirming that the high binding rate under the equality 
constraint is a feature of the constraint geometry rather than a 
learning artifact.

The wealth floor $W_t \ge W_{\min}$ is hit in less than $\floorHitRate$ 
of states, confirming that it acts as a safety guard rather than a 
perpetually active constraint. This aligns with the interpretation that 
the EZ policy uses intertemporal hedging to reshape the lower tail of 
the wealth distribution, reducing the probability of approaching the 
floor compared to myopic strategies.

\section{Conclusion and Future Work}

This paper takes a fairly standard long-run risk environment and asks a
simple question: what does intertemporal hedging demand look like once we
move to Epstein--Zin preferences and a realistically constrained
multi-asset setting, where closed-form solutions are no longer available?
To answer this, I set up a five-asset continuous-time model with a
persistent long-run risk factor, specify preferences using the usual set of
EZ parameters $(R,\psi,\delta)$, and solve the resulting
consumption--investment problem numerically with a projected
Pontryagin-guided deep policy optimization (P-PGDPO) scheme.

On the methodological side, the contribution is to show that P-PGDPO can be
extended in a fairly direct way from time-additive CRRA to EZ preferences,
provided that the EZ aggregator is written in a form compatible with the
continuous-time Hamiltonian. The value and costate processes are
approximated by neural networks, the policy is updated along the
Hamiltonian gradient, and portfolio and wealth constraints are enforced by
projection rather than by large penalty terms. This keeps the link to the
underlying stochastic control problem explicit and avoids some of the
pathologies that often appear in unconstrained reinforcement learning
setups.

On the economic side, the numerical results make the intertemporal hedging
motive visible in a concrete multi-asset example. The learned EZ policy
differs from the myopic benchmark in a way that is strongly state-dependent:
portfolio weights vary nonlinearly with both wealth and the long-run risk
factor, and the implied hedging component is sizable in regions where
investment opportunities are expected to move. At the same time, the
wealth constraint is respected in simulation, and the distribution of
terminal wealth is shifted away from the lower boundary, illustrating how
EZ preferences and long-run risk combine to shape dynamic hedging behavior.

There are several natural extensions that I do not pursue here. One is to
introduce additional sources of state variation, such as multiple
long-run risk factors or stochastic volatility, and to examine how the
structure of hedging demand changes as the dimension of the state space
increases. Another is to bring the model closer to data, either by
calibrating the LRR parameters to macro or asset-pricing evidence or by
embedding the P-PGDPO scheme in a simulated method-of-moments or
likelihood-style estimation framework. A third direction is to let the EZ
parameters themselves be learned or disciplined by external targets, rather
than fixed exogenously, and to study how the interaction between
risk aversion, intertemporal substitution, and constraints shows up in the
learned policy.

I view the present results as a step toward using continuous-time,
Pontryagin-guided deep learning tools as a complement to existing analytic
and numerical methods in dynamic portfolio choice with recursive
preferences. The combination of a structurally transparent control
formulation and flexible function approximation seems particularly useful
in settings where the underlying economics is reasonably well understood,
but exact solutions are out of reach once we move to multi-asset and
long-horizon environments.


\newpage
\appendix

\section{EZ Aggregator and the CRRA Limit}
\label{app:EZ-CRRA}

This appendix collects a simple observation about the continuous-time
Epstein--Zin aggregator used in Section~\ref{subsec:control-problem}.
The goal is just to make precise in what sense the specification in
\label{eq:EZ-aggregator} reduces to standard time-additive CRRA as
$\psi \to 1/R$.

\begin{proposition}
  Let $R>0$, $R\neq 1$ be fixed and let $\psi>0$, $\psi\neq 1$.
  Set $S = 1/\psi$ and $\theta = (1-R)/(1-S)$ and consider the
  Epstein--Zin aggregator
  \[
    f^{\mathrm{EZ}}(c,v)
    =
    \delta\, \theta\, v
    \biggl[
      \Bigl(
        \frac{c}{\bigl((1-R)v\bigr)^{1/(1-R)}}
      \Bigr)^{1-S}
      - 1
    \biggr].
  \]
  Suppose $v$ has the same sign as $1-R$ and define the CRRA
  utility $u(c) = c^{1-R}/(1-R)$.
  Then, for fixed $(c,v)$ in the admissible domain,
  \[
    \lim_{\psi \to 1/R} f^{\mathrm{EZ}}(c,v)
    = \delta \bigl( u(c) - v \bigr).
  \]
\end{proposition}

\begin{proof}[Proof (sketch)]
  Write $S = 1/\psi$ and observe that $S \to R$ as $\psi \to 1/R$.
  The key step is to expand the term
  $\bigl( c / ((1-R)v)^{1/(1-R)} \bigr)^{1-S}$ around $1-S = 1-R$,
  using a first-order Taylor expansion of the logarithm.
  This gives
  \[
    \Bigl(
      \frac{c}{((1-R)v)^{1/(1-R)}}
    \Bigr)^{1-S}
    =
    \frac{u(c)}{v} + o(1)
    \qquad\text{as } S \to R.
  \]
  Plugging this into the definition of $f^{\mathrm{EZ}}$ and taking
  the limit yields
  $f^{\mathrm{EZ}}(c,v) \to \delta(u(c)-v)$.
  A more detailed argument can be found in standard references on
  continuous-time Epstein--Zin preferences.
\end{proof}

\section{HJB, BSDE, and Pontryagin Conditions}
\label{app:HJB-BSDE-PMP}

For completeness, this appendix sketches how the HJB equation, the
BSDE representation, and the Pontryagin system fit together in the
continuous-time Epstein--Zin problem.

Starting from the BSDE
\[
  \dd V_t
  = - f^{\mathrm{EZ}}(c_t,V_t)\,\dd t
    + Z_t^\top \dd B_t,
  \qquad V_T = U(W_T),
\]
one can apply the Markovian ansatz $V_t = V(t,X_t)$ and use It\^o's
formula to obtain, formally,
\[
  \dd V_t
  = \bigl[ V_t + \mathcal{L}^u V(t,X_t)\bigr]\,\dd t
    + (\nabla_x V)^\top \sigma(X_t,u_t)\, \dd B_t,
\]
where $\mathcal{L}^u$ is the generator in \eqref{eq:generator}.
Matching the drift terms with the BSDE yields the HJB equation
\[
  0
  = V_t(t,x)
    + \sup_{u}
      \bigl\{
        f^{\mathrm{EZ}}(c,V(t,x))
        + \mathcal{L}^u V(t,x)
      \bigr\}.
\]

If $V$ is smooth and the optimizer $u^*(t,x)$ is unique, then the
Pontryagin system in
\eqref{eq:PMP-state-again}--\eqref{eq:PMP-stationarity-again}
can be obtained by setting
$\lambda_t = \nabla_x V(t,X_t^*)$ and differentiating the HJB
equation with respect to $x$ along the optimal trajectory.
Conversely, under appropriate regularity conditions, a solution of
the Pontryagin system gives rise to a candidate value function
$V(t,x)$ and policy $u^*(t,x)$ that solve the HJB.

The P-PGDPO algorithm in Section~\ref{sec:ppgdpo} effectively
works in the opposite direction: it parameterizes $V$ and
$\lambda$ directly, enforces approximate consistency with both the
BSDE and the gradient relation $\lambda \approx \nabla_x V$, and
then updates the policy to satisfy the Hamiltonian stationarity
condition.

\section{Alternative Definitions of Hedging Demand}
\label{app:alternative-hedging}

Throughout the main text, I define the intertemporal hedging component as the 
difference between the learned EZ policy and the myopic (Merton-style) benchmark:
\[
  \pi^{\mathrm{hedge}}(t,W,Y) 
  := \pi^{\mathrm{EZ}}(t,W,Y) - \pi^{\mathrm{myopic}}(t,W,Y).
\]
This definition is conceptually transparent and aligns with the classical 
decomposition in \citet{Merton1973}. However, alternative approaches exist 
in the literature, and it is useful to clarify why the current choice is 
appropriate for the present context.

\paragraph{State-hedging component (gradient-based).}
An alternative is to define hedging demand via the sensitivity of the portfolio 
to the state variable $Y$:
\[
  \pi^{\mathrm{state\text{-}hedge}}_i(t,W,Y) 
  := \frac{\partial \pi^{\mathrm{EZ}}_i(t,W,Y)}{\partial Y}.
\]
This measures how much the investor adjusts asset $i$ in response to changes 
in the LRR factor. A positive (negative) derivative indicates that the investor 
increases (decreases) exposure to asset $i$ when $Y$ rises, which can be 
interpreted as hedging against future shifts in the investment opportunity set.

\textit{Advantages:} This definition is model-free in the sense that it does 
not require specifying a myopic benchmark. It also aligns naturally with 
perturbation arguments in the stochastic control literature.

\textit{Limitations:} The gradient $\partial \pi_i / \partial Y$ depends on 
the units and scaling of $Y$, making cross-study comparisons less straightforward. 
Moreover, it does not directly quantify the \textit{magnitude} of hedging 
relative to the total portfolio, which is the primary object of interest in 
this paper.

\paragraph{Decomposition via backward induction.}
In discrete-time models, one can decompose the optimal policy by solving 
backward from the terminal date and isolating the myopic component at each 
step. This approach is common in the life-cycle consumption literature but 
is less natural in continuous time, where the HJB equation and the Pontryagin 
system provide the canonical characterization.

\paragraph{Conclusion.}
The $\pi^{\mathrm{EZ}} - \pi^{\mathrm{myopic}}$ decomposition used in this 
paper is the most direct way to isolate intertemporal motives in a setting 
where the myopic benchmark has a clear analytical interpretation. The main 
results (Figures~\ref{fig:asset1-surfaces} and~\ref{fig:hedging-vs-characteristics}) 
confirm that this hedging component is quantitatively significant and 
systematically related to the LRR factor exposure, as predicted by theory. 
Future work could complement this analysis by computing $\partial \pi / \partial Y$ 
via finite differences on the trained policy network, which would provide an 
additional diagnostic of state dependence.

\section{Additional Figures}
\label{app:additional-figures}

This appendix collects additional plots that are referred to in
the main text but omitted there for space reasons.

\begin{figure}[H]
  \centering
  \includegraphics[width=\textwidth]{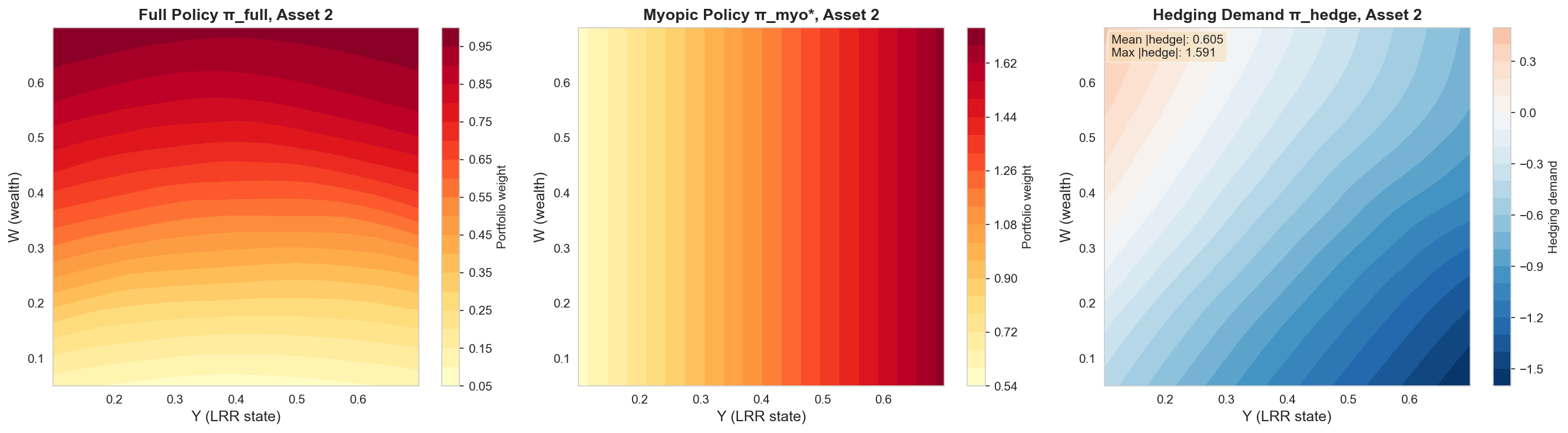}
  \caption{Portfolio, myopic benchmark, and hedging surfaces for
  Asset~2, analogous to Figure~\ref{fig:asset1-surfaces}.}
\end{figure}

\begin{figure}[H]
  \centering
  \includegraphics[width=\textwidth]{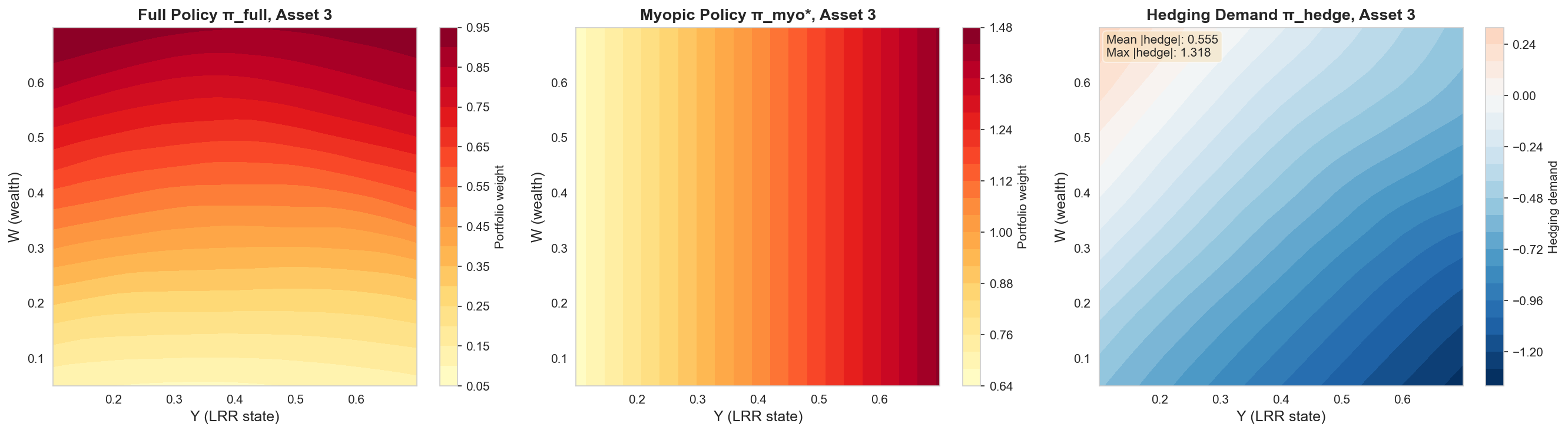}
  \caption{Portfolio, myopic benchmark, and hedging surfaces for
  Asset~3.}
\end{figure}

\begin{figure}[H]
  \centering
  \includegraphics[width=\textwidth]{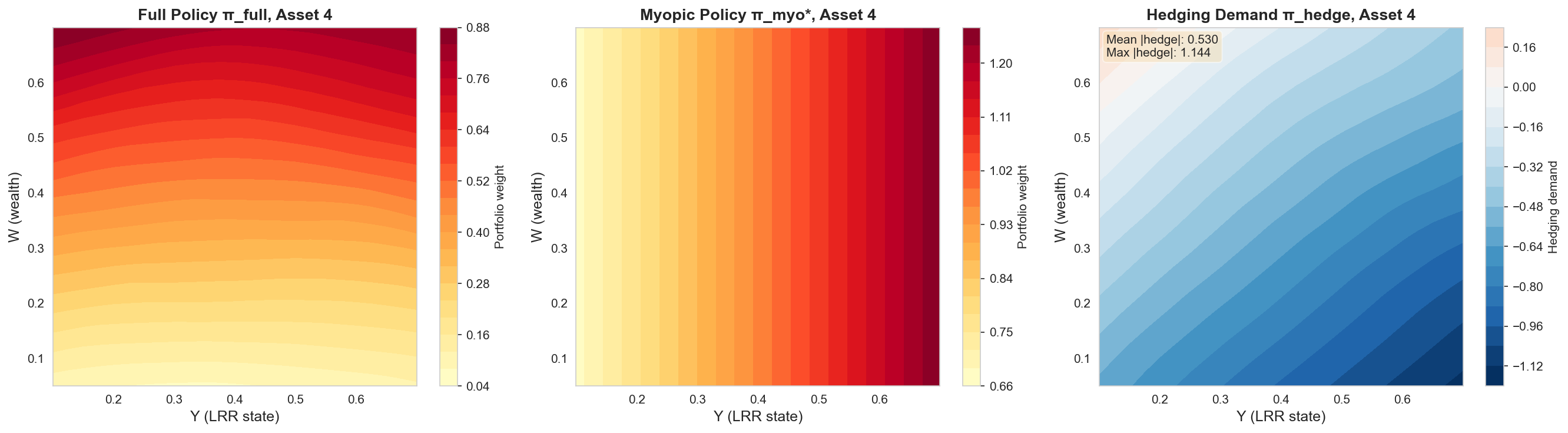}
  \caption{Portfolio, myopic benchmark, and hedging surfaces for
  Asset~4.}
\end{figure}

\begin{figure}[H]
  \centering
  \includegraphics[width=\textwidth]{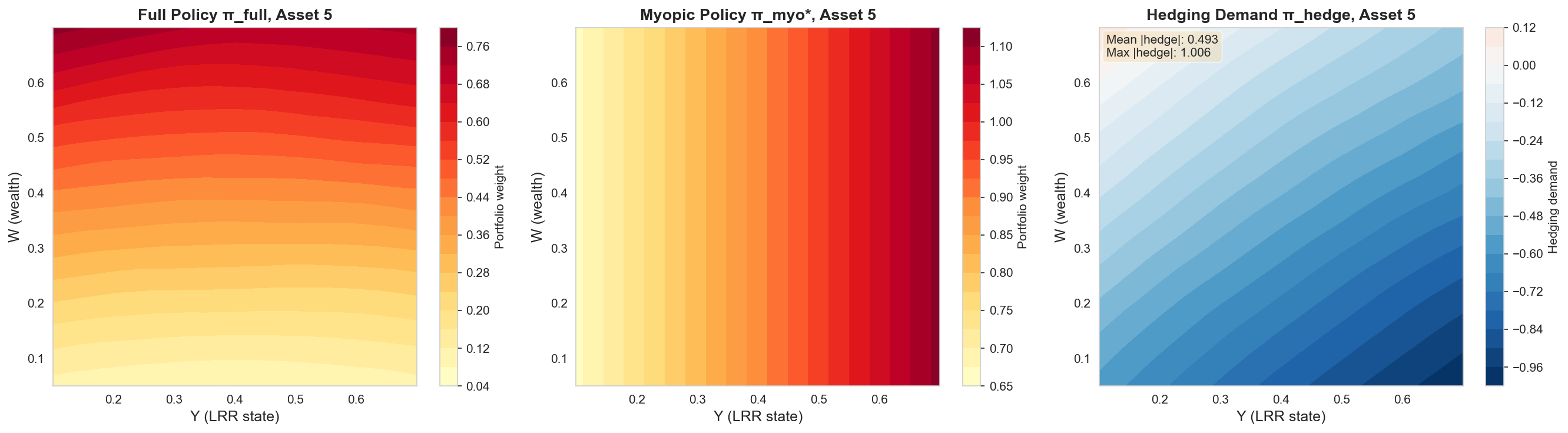}
  \caption{Portfolio, myopic benchmark, and hedging surfaces for
  Asset~5.}
\end{figure}

\end{document}